\documentclass[aps, pra, a4paper,showpacs,twocolumn,10pt]{revtex4-1}
\usepackage{bbm, amsmath, amssymb, amsthm, bm,textcomp, nicefrac,geometry,ragged2e}
\usepackage{graphicx,epstopdf,color,verbatim,enumitem}
\usepackage[dvipsnames]{xcolor}

\usepackage[bbgreekl]{mathbbol}
\usepackage{graphicx,epstopdf,color,verbatim,enumitem,ulem}
\usepackage[ruled]{algorithm2e}
\usepackage{algorithmic}

\definecolor{myurlcolor}{rgb}{0,0,0.7}
\definecolor{myrefcolor}{rgb}{0.8,0,0}
\usepackage[unicode=true,pdfusetitle, bookmarks=false,bookmarksnumbered=false,
bookmarksopen=false, breaklinks=false,pdfborder={0 0 0},backref=false,
colorlinks=true, linkcolor=myrefcolor,citecolor=myurlcolor,urlcolor=myurlcolor]
{hyperref}

\usepackage{pbox,hyperref,array}
\usepackage[caption=false]{subfig}

\newcommand{\ket}[1]{\left|#1\right\rangle}

\newcommand{\be}{\begin{equation}}
\newcommand{\ee}{\end{equation}}

\definecolor{darkorange}{RGB}{255,140,0}

\newcommand{\bea}{\begin{eqnarray}}
\newcommand{\eea}{\end{eqnarray}}

\def\k{{\bf k}}

\makeatletter
\newtheorem*{rep@theorem}{\rep@title}
\newcommand{\newreptheorem}[2]{%
\newenvironment{rep#1}[1]{%
 \def\rep@title{#2 \ref{##1}}%
 \begin{rep@theorem}}%
 {\end{rep@theorem}}}
\makeatother

\newreptheorem{theorem}{Theorem}

\newtheorem*{result*}{Result}

\newcommand{\px}{\sigma_x}

\newcommand{\pz}{\sigma_z}

\newcommand{\id}{id}

\def\ket#1{| #1 \rangle}

\def\dm#1{\left|#1 \right\rangle \left\langle #1 \right|}

\newtheorem{prop}{Proposition}\def\PRO{\begin{prop}}\def\ORP{\end{prop}}
\newtheorem{coro}{Corollary}\def\COR{\begin{coro}}\def\ROC{\end{coro}}
\newtheorem{theo}{Theorem}\def \TH{\begin{theo}}\def\HT{\end{theo}}
\def\TH{\begin{theo}}\def\HT{\end{theo}}
\newtheorem{defi}[prop]{Definition}\def\DE{\begin{defi}}\def\ED{\end{defi}}
\newtheorem{lemme}[prop]{Lemma}\def\LE{\begin{lemme}}\def\EL{\end{lemme}}

\DeclareGraphicsExtensions{.png,.pdf,.eps}

\begin{document}
\title{Modular architectures for quantum networks}
\author{A. Pirker}
\author{J.~Walln\"ofer}
\author{W.~D\"ur}
\affiliation{Institut f\"ur Theoretische Physik, Universit\"at Innsbruck, Technikerstr. 21a, A-6020 Innsbruck, Austria}

\date{\today}

\begin{abstract}
We consider the problem of generating multipartite entangled states in a quantum network upon request. We follow a top-down approach, where the required entanglement is initially present in the network in form of network states shared between network devices, and then manipulated in such a way that the desired target state is generated. This minimizes generation times, and allows for network structures that are in principle independent of physical links. We present a modular and flexible architecture, where a multi-layer network consists of devices of varying complexity, including quantum network routers, switches and clients, that share certain resource states. We concentrate on the generation of graph states among clients, which are resources for numerous distributed quantum tasks. We assume minimal functionality for clients, i.e. they do not participate in the complex and distributed generation process of the target state. We present architectures based on shared multipartite entangled Greenberger-Horne-Zeilinger (GHZ) states of different size, and fully connected decorated graph states respectively. We compare the features of these architectures to an approach that is based on bipartite entanglement, and identify advantages of the multipartite approach in terms of memory requirements and complexity of state manipulation. The architectures can handle parallel requests, and are designed in such a way that the network state can be dynamically extended if new clients or devices join the network. For generation or dynamical extension of the network states, we propose a quantum network configuration protocol, where entanglement purification is used to establish high-fidelity states. The latter also allows one to show that the entanglement generated among clients is private, i.e. the network is secure.

\end{abstract}

\maketitle


\section{Introduction}\label{sec:intro}

Quantum communication is an emerging discipline within quantum information science. The applications of quantum communication range from quantum key distribution \cite{ShorQKD,RennerPhd,ZhaoQKD,GottesmanQKD,LoQKD} over secure quantum channels \cite{SecChannelPortmann,SecChannelGarg,SecChannelBroadbent} to distributed quantum computation \cite{CiracDistributed} or quantum key agreement protocols \cite{XuQka,SunQka1,SunQka2}. All these tasks require entanglement as a key ingredient.
Quantum networks  \cite{Pirandola16, Meter2012,Meter2011,Schoute2016,Azuma2016a}, i.e. several connected quantum devices, form a highly relevant and active field of research in quantum communication. The most basic example of a quantum network corresponds to a single point-to-point quantum channel between two parties. 

In contrast to classical networks, where information is simply sent through a classical channel, quantum networks may serve a more general purpose due to the nature of quantum mechanics. In particular, sharing distributed entangled states between parties enables them to perform distributed quantum tasks. This includes e.g. distributed quantum computation, secure quantum channels in terms of teleportation \cite{BennettTele}, quantum conference key agreement and distribution \cite{XuQka,SunQka1,SunQka2,Epping2017,Chen07}, quantum secret sharing \cite{Chen07,Bell14,Markham08,Hillery99} or distributed sensing \cite{Komar14,Eldredge16}. Therefore it is desirable for a quantum network to offer capabilities beyond establishing pairwise communication channels, which corresponds to the quantum counterpart of classical network channels. Different multipartite entangled states serve as resources for various quantum task, but almost all of them rely on a class of states, so-called graph states \cite{He04}. Therefore, we concentrate on the generation of graph states in a quantum network in this work.

In contrast to previous approaches \cite{Campbell07,Matsuzaki10,Cuquet2012,Simon2006,Epping2016a}, we follow a top-down approach and introduce a flexible, modular architecture that allows one to establish arbitrary graph states among clients upon request, by using pre-generated (multipartite) entanglement. We aim at plug-and-play architectures, where clients solely connect to quantum network devices.
In addition we assume minimal functionality of clients, including only single qubit memories. In turn, there are no memory restrictions at the network devices, however we provide architectures that minimize the required storage qubits. We discuss and present solutions for quantum networks at an abstract level by identifying network building blocks and their interactions, which enables highly modular architectures. Moreover, by abstracting and defining network devices and networks as building blocks, our construction allows for layered and recursive structures.
The architecture we focus on is request-driven by the clients: they request a graph state from the network but do not participate in its complex construction process. As a consequence, the quantum network devices carry out the generation of the graph state in a collaborative and distributed manner. More precisely, the quantum network devices generate the desired target state from an entanglement resource state ---which we refer to as network state---, and a quantum state kept within the devices. The final state is then delivered to the clients. This enables a full multiparty functionality for the end users of the quantum network. For construction of the graph states we assume that the entanglement resource is available at the network devices upon incoming requests, i.e. the quantum state of the network has already been established. This has the advantage that no additional time is necessary to create the entanglement resource on demand. Hence, the time to generate the requested state only amounts to classical communication between devices, since all required measurements and operations can be done in a single time step. In addition, in such an entanglement-based approach, the network structure is at this stage in principle independent of the physical links.
Following such a top-down approach avoids complex routing tasks (see e.g. \cite{Leung2006, Epping2016b}) that need to be optimized dynamically while fulfilling the request. Observe that such tasks can cause a large time overhead which may result in long waiting times for the client. While the qubits for the network states need to be routed to their destination, the fact that these states are known in advance (due to the architecture of the network) and belong to a small set of states means that optimized routing schemes can be developed beforehand which do not need to be adjusted dynamically \footnote{Except in case of link failures on the routing paths, which can, in principle, also taken into account beforehand by alternative routing tables.}. Hence, the focus of quantum networks relying on our architectures lies on describing a straightforward way to distribute arbitrary graph states in the network and providing fast response times for its clients. This is not necessarily the most resource efficient way and if the state requested by the clients is known beforehand or the set of target states of the network is restricted, it would certainly be possible to develop optimized architectures for those applications.


Clients and quantum network devices may appear or also disappear dynamically during the lifetime of a quantum network. To take this into account, we propose a quantum network configuration protocol for distributing and extending the network state during the run-time of the network, which is crucial for enabling our approaches. Observe that the entanglement resource needs to be such that the network devices can generate {\it any} graph state from it. 
Furthermore, it turns out that our protocols are especially suited for parallelization within the network. By parallelization we mean here that disjoint subsets of clients may demand the network devices to generate graph states among these subsets in parallel, without redistributing a new network state. Our architectures follow a recursive approach, where networks and devices appear in a recursive manner. This turns it into a promising candidate for recursive quantum network architectures \cite{Meter2011}.

To establish states with high fidelity among the devices, we rely on encoded transmission using quantum error correction, or bipartite and multipartite repeater architectures. Both approaches allow one in principle to establish long-distance entanglement in a network. The latter includes entanglement distillation, where few high fidelity states are generated from multiple noisy copies. Here, we employ entanglement distillation also to show the security of our architectures. In particular, as shown in  \cite{Zwerger17,Pirker16a}, it suffices to prepend entanglement distillation protocols appropriately to disentangle any eavesdropper. This generates private entanglement, and yields to secure quantum networks even in cases where network devices are untrusted. Finally we point out that if the functionality of a network is restricted to generate specific classes of states, like e.g. GHZ states, our architectures simplify tremendously and provide efficient solutions for quantum networks.

The paper is organized as follows. In Sec. \ref{sec:back} we review the basic concepts which will be used throughout this paper. In particular we discuss GHZ states, graph states, their manipulations and merging/connecting techniques, generation of high-fidelity states over large distances and prior work on quantum networks. Then we present in Sec. \ref{sec:arch} the elementary building blocks of quantum networks. Next, we propose the network architecture which relies solely on GHZ states in Sec. \ref{sec:arch:ghz} and the decorated network architecture in Sec. \ref{sec:arch:universal}. We also present a hybrid architecture in Sec. \ref{sec:arch:hybrid} and discuss an architecture relying solely on bipartite entangled states in Sec. \ref{sec:arch:bellpairs}. Next, we discuss protocols for generating and extending the network state of the architectures of Sec. \ref{sec:arch:ghz} and \ref{sec:arch:universal} in Sec. \ref{sec:qncp}. In Sec. \ref{sec:compare} we compare and relate our architectures to the bipartite approach of Sec. \ref{sec:arch:bellpairs}. Finally, we discuss optimizations of our architectures in Sec. \ref{sec:optimized}. We conclude and summarize our findings in Sec. \ref{sec:conclusion} where we also provide an outlook and identify interesting open questions.

\section{Background}\label{sec:back}

In the following we summarize the necessary background to construct quantum networks. In particular we start with a short review of graph states, a specific class of stabilizer states, and their manipulations. Then we recap some merging and connecting techniques related to graph states as well as basic and useful properties of GHZ states. Next, we recall briefly the basics of entanglement distillation protocols and quantum repeaters. Finally, we review some relevant results on quantum networks.

\subsection{Graph states and their manipulations}\label{sec:back:meas}

Graph states \cite{He04,Guehne05,He06,Toth06} form a specific class of stabilizer states. A stabilizer state is a quantum state which is stabilized by operators of the Pauli group. In particular, for a stabilizer state $\ket{\psi}$ there exists a subgroup $S_{n}$ of the Pauli group $P_{n}$ such that we have for all $s \in S_{n}$ that $s \ket{\psi} = \ket{\psi}$. In other words, the state $\ket{\psi}$ is the common $+1$ eigenstate of the operators within $S_{n}$. A graph state $\ket{G}$ \cite{He06} associated with a classical graph $G=(V,E)$ is defined as the common eigenstate of the stabilizers or correlation operators
\begin{align}
K_{a} = X_{a} \prod_{\lbrace a,b \rbrace \in E} Z_{b} \label{eq:graphstate:correlation}
\end{align}
where $a \in V$, and $E$ denotes the set of edges. The subscripts in (\ref{eq:graphstate:correlation}) indicate on which qubit the Pauli operator acts on. 

Graph states can explicitly be generated by means of controlled phase gates among all edges. More specifically, the graph state $\ket{G}$ stabilized by the family of operators (\ref{eq:graphstate:correlation}) can be generated via
\begin{align}
\ket{G} = \prod_{\lbrace a,b \rbrace \in E} CZ_{a,b} \ket{+}^{\otimes V}
\end{align}
where $CZ_{a,b} = \dm{0}_{a} \otimes \id_{b} + \dm{1}_{a} \otimes Z_{b}$ denotes the controlled phase gate, also referred to as CZ gate. In particular, performing a controlled phase gate between two non-adjacent qubits introduces a new edge within the graph $G$. Conversely, if there is already an edge within the graph $G$ the controlled phase gate between the two adjacent qubits will remove that edge from the graph. 

There are several important properties of graph states \cite{He04,He06} which we make use of extensively here. For example, if a qubit is measured in the $Z$ basis, then all edges associated with this qubit and the vertex corresponding to that qubit will be deleted from $G$. Depending on the measurement outcome, local Clifford corrections may need to be applied to the remaining qubits. If a qubit is measured within the $Y$ basis, the subgraph spanned by the neighbours of the measured qubit is locally complemented. After that the qubit is removed from the graph. Again, depending on the measurement outcome correction operations need to be applied. Finally, if a qubit is measured within the $X$ basis, one selects a neighbour, which we denote by $b$, of the measured qubit, performs a local complementation w.r.t. $b$ and locally complements the resulting graph w.r.t. the measured qubit. Next the measured qubit is removed from the graph. As last step another local complementation w.r.t. $b$ is done. Depending on the measurement outcome, correction operations need to be employed.

Furthermore we call two graph states $\ket{G}$ and $\ket{G'}$ local unitary equivalent (LU equivalent), if there exist unitaries $U_1, \ldots, U_n$ such that $\ket{G'} = U_1 \otimes \cdot \otimes U_n \ket{G}$.

\subsection{Connecting and merging graph states}\label{sec:back:merge}

In this section we will describe how to connect and merge graph states. Since the goal of our architectures is to generate arbitrary graph states, the techniques which we discuss here are crucial when applying our protocols. For simplicity we do not concern ourselves with any Clifford corrections due to measurement outcomes at this point.

\subsubsection{Connecting graph states}\label{sec:back:merge:czconnect}

\begin{figure}[h!]
\begin{center}
\scalebox{4}{
\includegraphics{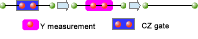}
}
\caption[h!]{\label{fig:czconnect} Graphical illustration of the connecting procedure.}
\end{center}
\end{figure}

The first technique, which we also refer to as {\it connecting procedure}, uses controlled phase gates and measurements in the $Y$ basis. Suppose we want to connect two graph states at vertices having only one neighbour each. Applying a CZ gate between the those vertices establishes an edge between them. Recall that a measurement in the $Y$ basis performs a local complementation of the subgraph induced by the neighbourhood of the measured vertex. Since both vertices which have been connected by the CZ gate had one neighbour before the CZ gate each, measuring those vertices results in a wire in the final graph state, see Fig. \ref{fig:czconnect}.

%
%

\subsubsection{Merging graph states}\label{sec:back:merge:cnot}

\begin{figure}[h!]
\begin{center}
\scalebox{6}{
\includegraphics{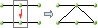}
}
\caption[h!]{\label{fig:universal:cnot} Transformation of a tensor product of two graph states under a controlled NOT.}
\end{center}
\end{figure}

Finally we discuss a merging procedure for graph states which uses a controlled NOT (CNOT) gate. We recall the effect of a CNOT on a tensor product of two graph states $\ket{G_1} \otimes \ket{G_2}$ where $G_1=(V_1,E_1)$ and $G_2=(V_2,E_2)$ denote classical graphs. For that purpose we first compute the commutation relations of a CNOT with $X$ and $Z$ operators as
\begin{align}
\mathrm{CNOT}_{s \to t} X_s = \left(X_s \otimes X_t \right) \mathrm{CNOT}_{s \to t}, \label{eq:cnot:1} \\
\mathrm{CNOT}_{s \to t} Z_s = Z_s \mathrm{CNOT}_{s \to t}, \label{eq:cnot:2} \\
\mathrm{CNOT}_{s \to t} X_t = X_t \mathrm{CNOT}_{s \to t} \label{eq:cnot:3}, \\
\mathrm{CNOT}_{s \to t} Z_t = \left(Z_s \otimes Z_t \right) \mathrm{CNOT}_{s \to t} \label{eq:cnot:4}
\end{align}
where $s$ denotes the source and $t$ denotes the target qubit of the CNOT. Therefore we find for the application of a CNOT to the tensor product of the two graph states $\ket{G_1}$ and $\ket{G_2}$ where $s \in V_1$ and $t \in V_2$ the following: Eq. (\ref{eq:cnot:2}) implies that all stabilizers of the neighbourhood of $s$ do not change, since $Z_s$ commutes with $\mathrm{CNOT}_{s \to t}$. Similarly we observe that the same observation applies to the stabilizer of the target qubit according to (\ref{eq:cnot:3}), since $X_t$ commutes with $\mathrm{CNOT}_{s \to t}$. In contrast, (\ref{eq:cnot:1}) implies that all neighbours of the target qubit $t$ will be neighbours of the source qubit $s$ after the CNOT, since we annihilate the $X_t$ operator appearing within $K_s$ after the CNOT due to (\ref{eq:cnot:1}) by multiplying $K_s$ and $K_t$, thereby producing a new stabilizer. This transformation of the neighbourhood of the source qubit $s$ is also apparent from (\ref{eq:cnot:4}). So in summary, applying a CNOT to the tensor product of two graph state introduces new edges in the resulting graph state between the source qubit and the neighbourhood of the target qubit, see Fig. \ref{fig:universal:cnot} and \cite{EPPallGraphs}. 

Therefore, if one measures the target qubit of the CNOT in the $Z$ basis one effectively moves the neighbourhood of the target qubit to the source qubit of the CNOT. 

Observe that the same transformation rule applies to a single graph state where the source and target qubit of the CNOT are not adjacent and have disjoint neighbourhoods.

\subsection{GHZ states}\label{sec:back:ghz}

GHZ states are of particular interest in quantum information science, as they can be used to generate bipartite entanglement among any pair of qubits, are genuinely non-local and genuinely multiparty entangled. These states have several important applications like e.g. within quantum key agreement protocols \cite{XuQka}, quantum conference key distribution \cite{Epping2017}, cryptographic protocols in general \cite{Chen07}, quantum secret sharing \cite{Hillery99} or quantum metrology \cite{Geza14,Dooley16}. In \cite{Epping2017} it has been shown that a multipartite approach to quantum conference key distribution outperforms bipartite strategies in networks with a bottleneck.  

A $n$-qubit GHZ state is of the form
\begin{align}
\ket{\mathrm{GHZ}_{n}} = \frac{1}{\sqrt{2}}\left(\ket{0}^{\otimes n} + \ket{1}^{\otimes n}\right). \label{eq:ghz}
\end{align}
The following observation forms one of the key ingredients within this paper: Suppose we connect two GHZ states $\ket{\mathrm{GHZ}_{m}}$ and $\ket{\mathrm{GHZ}_{n}}$ via a Bell-measurement, which is a two qubit projective measurement with projectors $\dm{B_{ij}}$ where $\ket{B_{ij}} = (\id \otimes \px^j \pz^i) (\ket{00} + \ket{11})/\sqrt{2}$ and $i,j \in \lbrace 0,1 \rbrace$ denotes the four Bell-basis state projectors. Then the probability for each outcome is exactly $1/4$. More importantly, the post-measurement state is, up to local Pauli corrections, again a GHZ state, i.e. the post-measurement state after applying the correction is $\ket{\mathrm{GHZ}_{m+n-2}}$. In addition, there also exists a measurement where two qubits of the GHZ states $\ket{\mathrm{GHZ}_{m}}$ and $\ket{\mathrm{GHZ}_{n}}$ merge into one qubit, resulting in the GHZ state $\ket{\mathrm{GHZ}_{n+m-1}}$. This is accomplished via measuring w.r.t. $|0 \rangle \langle 00| + |1 \rangle \langle 11|$ and $|0 \rangle \langle 01| + |1 \rangle \langle 10|$.   

These observations have been used in \cite{Wallnofer16_2D} to extend the quantum repeater scheme to the two-dimensional case, i.e. a quantum repeater for GHZ states. Our architectures will also rely on this result to cope with large distances between network devices within a quantum network. 

\begin{figure}[h!]
\begin{center}
\scalebox{4}{
\includegraphics{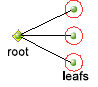}
}
\caption[h!]{\label{fig:ghz} Graph state that is LU equivalent to a four qubit GHZ state. The state is related by Hadamard operations (indicated by red circles) on all leaf qubits to a standard GHZ state, see text for details. The remaining qubit is denoted as root qubit of the GHZ state.}
\end{center}
\end{figure}

Another observation is that the $n$-qubit GHZ state is LU equivalent to a graph state. To see this, recall that a graph state is stabilized by operators of the form (\ref{eq:graphstate:correlation}). One easily verifies that the GHZ state (\ref{eq:ghz}) is stabilized by the operators $X^{\otimes n}$ and $Z_1 \otimes Z_i$ where $2 \leq i \leq n$. Now suppose we apply a Hadamard rotation to all qubits excepts the first. Since stabilizers transform via conjugation and $H X H = Z$ as well as $H Z H = X$, we find that the stabilizers of the resulting state after applying the Hadamard rotation are given by $X_1 \otimes Z^{\otimes n-1}$ and $Z_1 \otimes X_i$ for $2 \leq i \leq n$. In this paper we often use the graphical representation of this graph state (see Fig. \ref{fig:ghz}) to indicate a GHZ state. In particular, we call qubit $1$ of this graph representation the \textit{root} and qubits $2, \ldots, n$ the \textit{leafs} of the GHZ state.

\subsection{Quantum communication over noisy channels: quantum repeaters and entanglement distillation}
\label{sec:longdistance}
In order to establish quantum networks, there are mainly three different strategies: direct transmission of quantum states over quantum channels, networks relying on bipartite entanglement and networks relying on multipartite entanglement. The direct transmission of quantum states corresponds to sending a qubit directly through a quantum channel, e.g. an optical fiber or free space, to a recipient. Unfortunately quantum information is very fragile against decoherence and channel loss, and the direct transmission of a qubit is limited to rather short distances. The classical approach to restore the signal periodically after a certain distance using classical repeaters cannot be straightforwardly adapted because of the no-cloning theorem \cite{Wootters82}, which states that quantum information cannot be simply copied or amplified. The fundamental limits of repeaterless point-to-point quantum communications where derived in \cite{Pirandola2017}. However, encoding the qubit to be sent using a quantum error correction code \cite{Nielsen} turns out to be useful for protecting quantum information against errors introduced by a quantum channel. 

An alternative strategy is to establish bipartite entangled states using a so-called quantum repeater scheme \cite{Br98,DurRepeater,Sa09}. A quantum repeater is able to establish a long-distance Bell-pair by combining several short-distance Bell-pairs via entanglement swapping. A perfect Bell-pair is equivalent to a secure quantum channel \cite{Hayden,SecChannelBroadbent,SecChannelPortmann,SecChannelGarg} by means of teleportation \cite{BennettTele}. The original quantum repeater scheme \cite{Br98,DurRepeater} executes entanglement distillation and entanglement swapping in a nested fashion, thereby generating high-fidelity entangled pairs over larger and larger distances. This allows one to obtain an efficient scheme to generate long-distance entangled pairs, overcoming the exponential scaling of resources (time or channel usages) of direct transmission. Recently, a generalized repeater scheme to generate multipartite entangled GHZ states has been put forward \cite{Wallnofer16_2D}. Quantum repeaters have been extensively used and discussed also in the context of quantum networks, see e.g. \cite{Munro15,Munro08,Sa11,Guha15,Pant17a,Das17,Pant17,Meter2013a,Meter2013b,VanMeter2014book,Meter2009,Munro2010,Muralidharan2016,Epping2016b,Hayashi2007}.

A central element of quantum repeaters are (bipartite) entanglement distillation protocols (sometimes also referred to as entanglement purification protocols) that aim at distilling a perfect Bell-pair via local operations and classical communication (LOCC) from several noisy copies of a Bell-pair. Depending on the mode of operation we distinguish between recurrence-type \cite{BennettRecurrence,Deutsch} and hashing protocols \cite{Be96}. There also exist extensions of entanglement distillation protocols to multipartite settings \cite{Du05,EPPallGraphs,Duer07,Miyake05,Fujii09,Chen07,ManevaEPP,HostensEPPGraph,GlancyEPPGraph,HostensEPPGraph2}, in particular for all graph states \cite{EPPallGraphs,Duer07}. Notice that entanglement distillation protocols are simpler for two-colorable graphs (i.e. where the vertices can be grouped into two disjoint sets, where there are no edges within each group). It is hence desirable to design architectures that rely solely on two-colorable graph states, as they can be more efficiently purified in noisy settings.




\subsection{Quantum networks}

Quantum networks constitute an infrastructure which aims at connecting several distant quantum clients. The applications range from point-to-point teleportation, distributed quantum computation, quantum key distribution, quantum secret sharing to quantum key agreement, see Sec. \ref{sec:intro}. 

In the following we highlight a few papers that are relevant in our context. The application of the quantum repeater to long-distance quantum networks has received tremendous attention in recent years, especially the system design and architecture of quantum repeater network \cite{Pirandola16,Meter2013a,Meter2013b,Meter2009,Munro2010,Muralidharan2016}. In \cite{Pirandola16} the tightest upper bounds for quantum communication, entanglement distribution and key generation over an arbitrary quantum network using single-path or multi-path routing protocols was derived. The design of quantum repeater networks was investigated in \cite{Meter2013a}. In \cite{Meter2013b} the path selection within a quantum repeater network based on the Dijkstra algorithm according to certain metrics was investigated. In addition, a layer model similar to the Open Systems Interconnection Model (OSI model) \cite{Zimmermann88} for classical computers was proposed in \cite{Meter2012,Meter2011,Meter2009}. These layer models include layers for entanglement distillation, entanglement control and link management. In principle, bipartite quantum repeaters suffice to construct any given state within a quantum network. The construction of explicit quantum repeater protocols was investigated in \cite{Azuma2016}. Protocols for distributing bipartite entanglement with quantum repeaters were studied in \cite{VanDam2017, Jones2016}. Another line of investigation pursues the application of quantum network coding techniques to quantum repeaters \cite{Epping2016b,Hayashi2007}. However, observe that in order to generate arbitrary graph states within a network relying on Bell-pairs, the repeater nodes still need to combine the Bell-pairs according to the requested graph state. 

Security aspects of direct transmission of quantum information within a quantum network was studied in \cite{Sun2016}. This proposal does neither rely on quantum repeaters nor on any entanglement, but a-priori QKD links were mandatory to achieve security. 
One approach to design scalable quantum networks is the quantum recursive network architecture (QRNA) \cite{Meter2011}. There the quantum network is composed of elementary building blocks like quantum repeaters or networks in a recursive fashion. Security aspects within QRNA-based architectures have recently been studied in \cite{Amoretti2017} where the authors investigate a verification protocol to detect an eavesdropper. However, current descriptions of QRNA still reside within a high level of abstraction and rely on bipartite entanglement via quantum repeaters rather than multi qubit quantum states. 

Nevertheless, the authors of \cite{Meter2011} already identify one key aspect of quantum networks which we also believe is of utmost importance: generating/establishing distributed quantum states among the network without direct interaction of the clients, see also Sec. \ref{sec:intro}. In particular, one important aspect of quantum networks is to generate certain distributed states within a given network on demand. For example, several clients of a quantum network may wish to share a GHZ state at time $t$ in order to agree on a conference key. At time $t+1$ two clients within the same network may wish to establish a secure quantum channel. This implies that the network needs to offer capabilities that allow to generate different target states between clients on demand, at any time and in an ideal case without direct interaction between the clients. Here we concentrate on the generation of arbitrary graph states $\ket{G}$. Notice that this includes also states corresponding to parallel requests of disjoint parties, as a tensor product of small graph states is again a graph state of $N$ parties. 

Ideally the network should respond immediately to such requests. This can only be accomplished in a setting with pre-distributed entanglement resources which the network uses for graph state generation. Observe that this corresponds to a top-down approach, where this "universal" entanglement resource is available prior to requests. The inverse direction, i.e. generating a graph state by fusing small scale entangled states, was studied e.g. in \cite{Campbell07} and \cite{Matsuzaki10}. More precisely, in \cite{Campbell07} adaptive strategies for the growth of graph state in the presence of monitored errors were presented. An approach for probabilistic growth of graph states by the fusion of small elementary graph states was proposed in \cite{Matsuzaki10}. 

The task of generating graph states was addressed in \cite{Cuquet2012} for a single quantum network. Two of the protocols the authors propose rely on a central master node which prepares the requested graph state locally and teleports it to the clients. However, these protocols require {\it one} central node within the network which performs the local preparation of the target state and it remains unclear how several such central nodes connect and collaborate. The third protocol they propose uses GHZ states, which is interesting and promising, but requires explicit knowledge of the final graph state and the application of CZ gates and measurements in the $Y$ basis at the clients. Therefore the clients need to be able to apply two qubit entangling gates. In \cite{Simon2006} it was studied how to establish an arbitrary graph state within a network via broker qubits. There, the entanglement structure is first established among the broker qubits, which is assumed to be error-prone, and then projected onto the client qubits. Finally, \cite{Epping2016a} uses the quantum repeater scheme and quantum error correction to establish large-distance graph states. 

Most of the previous works consider a bottom-up approach to quantum networks, where resources are generated on demand. Therefore, prior to completing a particular task, the network needs to distribute the necessary resources. So the network devices need to perform routing tasks in order to determine a path for resources through the network. As we explained in Sec. \ref{sec:intro} this introduces longer waiting times for the clients of a network and we eliminate this issue by following a top-down approach to quantum networks. 

\subsection{Our setting}

We recall our setting as explained in Sec. \ref{sec:intro}: We consider the scenario where an entanglement resource is distributed in a quantum network prior to any request. Furthermore the resource -- which we also refer to as network state -- is such that any arbitrary graph state can be generated from it.  This corresponds to a top-down approach, where the network devices manipulate the entanglement resource in a collaborative and distributed manner to fulfill requests in a network. The clients of the network remain passive throughout the entire process, i.e. they only request graph states rather than constructing them.   

The protocols we propose here have significant advantages compared to earlier works. First, they explicitly use genuine multipartite entangled quantum states which is suited to the task of distributing multipartite entanglement over the network. Second, the control of which edge in the target graph state will be established reside within the network devices, which is in contrast to earlier works \cite{Cuquet2012,Simon2006,Epping2016a}. The clients themselves only need to store a small number of qubits and do not need the ability to perform two qubit measurements and entangling gates. The network devices need the capability to employ two qubit entangling gates and Bell-measurements. In addition, we remove the necessity of a single central master node within a network as in \cite{Cuquet2012}. The network devices we propose collaborate and generate the target graph state in a distributed, collective manner, i.e. we follow a top-down approach where the network devices consume the entanglement resource available prior to requests. This also minimizes the waiting times for clients and allows for the parallel generation of graph states. In addition our approach pursues a plug-and-play architecture, similar to computer networks, which offers high modularity at the network level. In particular only small changes are required for modification of the network, i.e. when adding or removing network elements or clients. 

As we discussed in Sec. \ref{sec:intro}, the architectures we study can be layered due to its modular design. Therefore, a quantum network may be part of a larger quantum network without modifying the basic building blocks we propose. We remark that one can also introduce new network layers on demand, which can help to increase connectivity or reduce the complexity to establish certain target states among clients. In this way one can overcome limitations given by network structure at the level of physical links, and e.g. introduce a new layer to provide a shortcut between nodes that are otherwise connected only indirectly via multiple network elements.
Finally, the computational assumptions we apply to our clients are minimal, i.e. the clients can apply single qubit gates and single qubit measurements only.

\section{Elementary building blocks for quantum networks}\label{sec:arch}

In this section we describe three different types of elementary building blocks for quantum networks: quantum clients, quantum graph state switches and quantum graph state routers. The latter two constitute the network devices we consider. If the context is clear we will also refer to those entities as clients, switches and routers. In addition we also refer to switches and routers as (network) devices.

We remark that these devices and their functionality are in analogy to classical devices, however due to their quantum nature they need to have additional features. Notice that not all classical devices can have a natural quantum analogue. Consider for instance a classical hub that distributes the same information to several other devices. If information is unknown, a quantum hub is not possible as unknown quantum states can not be copied. We hence do not consider hubs here. 



\subsection{Quantum clients}\label{sec:arch:build:client}

Quantum clients correspond to the clients in a quantum network who wish to share arbitrary graph states on requests. 

\begin{figure}[h!]
\begin{center}
\scalebox{2.5}{
\includegraphics{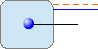}
}
\caption[h!]{\label{fig:qclient} The figure depicts a quantum client: internally it stores one qubit which might be a half of a Bell-pair to a network device. The blue line indicates a quantum channel, the orange dashed line a classical authenticated channel and the black line entanglement. A quantum client has the capability to apply single qubit unitaries in our very basic setup.}
\end{center}
\end{figure}

We assume that quantum clients solely connect to network devices via a quantum channel, e.g. an optical fibre, see Fig. \ref{fig:qclient}, or that they share entanglement with a network device. In particular we do not assume any kind of quantum channel between two quantum clients. We further constrain the clients to neither have any knowledge about the topology of the quantum channels within the network nor to have the capabilities to perform any two qubit quantum operation or measurement. This constraints imply that the network devices need to handle the generation process of the target graph states. This is in contrast to previous works \cite{Simon2006,Cuquet2012} where the clients were responsible for generating the target graph state. \newline
We summarize the capabilities of a quantum client as follows: only single qubit gates are feasible and the clients do not participate in the construction mechanism of the requested graph states. This implies that the clients need to store only a single qubit. Therefore the clients, or end-users, have minimal functionality.

We remark that to ensure security in a completely untrusted network, the clients will additionally need the capability to participate in an entanglement distillation protocol, which is discussed in more detail in section \ref{sec:arch:ghz:sec}, thereby implying that the clients need the ability to perform entangling gates and Bell-measurements in this scenario. 

\subsection{Quantum graph state switches}\label{sec:arch:build:switch}

Quantum graph state switches form the quantum pendant to classical switches, which operate on the so-called "data link layer" of the OSI model \cite{Zimmermann88} in computer networks. A classical switch has the capability to separate broadcast domains and perform a basic routing by using hardware addresses of network interfaces. Quantum graph state switches provide several interfaces to connect with the quantum network, e.g. through optical fibres. 

\begin{figure}[h!]
\begin{center}
\scalebox{4}{
\includegraphics{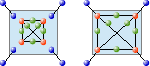}
}
\caption[h!]{\label{fig:ghz:black} In the left figure each quantum client shares a Bell-pair with the quantum graph state switch. The quantum graph state switch uses this Bell-pair to finally teleport the graph state to the quantum clients. Alternatively a switch might directly connect the client qubit to its device state see the right figure. This approach saves storage space at the network device.}
\end{center}
\end{figure}

Quantum clients connect to switches via the interfaces of the switch. In doing so the quantum graph state switch sends one qubit to the client. This qubit might correspond to half of a Bell-pair. In that case the other half resides within the quantum graph state switch, see Fig. \ref{fig:ghz:black}. But a switch might use a different mechanism, relying on multi-qubit entangled states and measurements in the $X$ or $Y$ basis, to connect clients, see Sec. \ref{sec:arch:ghz:device} and Sec. \ref{sec:arch:universal:device}. Furthermore, switches might also connect to other network devices within a quantum network via the same interfaces, in a similar fashion as in classical computer networks. 

Suppose a quantum client requests a particular graph state from the network. The quantum graph state switch is responsible for generating the adjacency of the clients which it connects. Observe that such adjacencies might also involve clients located at other network devices. In order to fulfill this task, the switch uses internally a quantum state, which we also refer to as {\it device state}. 

We determine the demands on such a state easily: First, the switch needs to be able to establish any graph state between its connected clients from this resource, and second, the switch needs the capability to generate adjacencies across the network to clients located at other network devices. In addition we assume that the device state of a switch is available prior to a request. Hence the switch needs to generate the adjacency of a request by manipulating that resource appropriately. From this we deduce that switches need to store and manipulate qubits, which means that they use quantum memory, apply single and two qubit unitaries and perform measurements. 

We will propose two different device states in Sec. \ref{sec:arch:ghz:device} and Sec. \ref{sec:arch:universal:device}. 

\subsection{Quantum graph state routers}\label{sec:arch:build:router}

Quantum graph state routers are the most involved network devices within quantum networks, as they might connect several quantum networks. Their responsibilities are the generation of the network state (if the router is dedicated for doing so), which we refer to as {\it quantum network configuration protocol (QNCP)} server, see Sec. \ref{sec:arch:build:QNCP}, and the ability to connect two or possibly more quantum networks. The term QNCP server originates from the classical DHCP ({\it dynamic host configuration protocol}) server, which provides the logical addresses, i.e. the IP addresses, in computer networks. In addition, a quantum graph state router offers the same functionality as a switch, see Sec. \ref{sec:arch:build:switch}. 

Furthermore, routers may also provide a classical service to network devices: They have a global view on the network, i.e. they know which client connects to which network device of a network. This information is crucial if the network has to generate specific graph states, see Sec. \ref{sec:arch:build:network} for a more detailed discussion. 

Quantum graph state routers correspond to classical routers, which operate on the third layer of the OSI model (network layer), since they are able to connect networks, provide logical addresses and perform routing tasks.

\subsection{Quantum networks}\label{sec:arch:build:network}

\begin{figure}[h!]
\begin{center}
\scalebox{3}{
\includegraphics{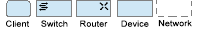}
}
\caption[h!]{\label{fig:basic:legend} The different symbols we use for representing clients, switches, router, devices and networks.}
\end{center}
\end{figure}

In the following we use the graphical representation of clients, switches, routers, devices (which are switches or routers) and networks depicted in Fig. \ref{fig:basic:legend}. 

\begin{figure}[h!]
\begin{center}
\scalebox{2.5}{
\includegraphics{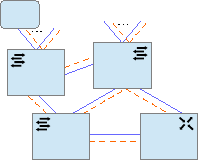}
}
\caption[h!]{\label{fig:basic:network} Illustration of a basic hardware setup. The clients connect to network devices, e.g. switches or routers through a quantum channels (blue line) and a classical authenticated channel (orange dashed line). Network devices also connect via quantum channels and classical authenticated channels.}
\end{center}
\end{figure}

\begin{figure}[h!]
\begin{center}
\scalebox{2.5}{
\includegraphics{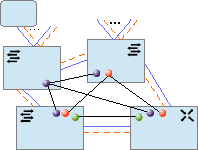}
}
\caption[h!]{\label{fig:basic:network:ent} In this example the physical channel configuration (blue edges -- quantum channels, orange dashed edges -- classical channels) differs from the entanglement structure of the network state. The vertices correspond to qubits of the network state and the black edges depict entanglement.}
\end{center}
\end{figure}

We pursue an entanglement-based approach to quantum networks via multi-partite states. More precisely, the network devices of a network connect at a physical level via quantum channels like e.g. optical fibres, see Fig. \ref{fig:basic:network}.
The network devices use these channels to distribute highly entangled states between each other, which we refer to as {\it network state}. Using this state a device is able to generate adjacencies of his clients to clients connected to other network devices, i.e. adjacencies across its device boundary. We emphasize that the entanglement structure of the network state may be completely different from the channel configuration of the network, see Fig. \ref{fig:basic:network:ent}. Furthermore, if network devices are far distant, we employ the techniques discussed in Sec. \ref{sec:longdistance} to establish a long-distance network state. Observe that the same techniques can be applied for far-distant clients of a network device.

Generating target graph states via network states has many advantages. First of all it is fast, since the network state is distributed prior to requests. Therefore no distribution times need to be taken into account on incoming requests, which is in contrast to earlier proposals. The time clients have to wait for the graph state simply amount to the classical communication time, since the quantum operations and correction operations can be done within a single time step. Furthermore this approach allows for full flexibility, as the network devices tailor the network state according to incoming requests. As we will see in Sec. \ref{sec:arch:ghz:net} and Sec. \ref{sec:arch:universal:net} our network states also enables the parallel generation of graph states on disjoint subsets of clients. 

The network state is, as the device state, assumed to be a static resource which the devices consume during graph state generation. Therefore we imply a similar constraint as for the device state also to the network state: using the device {\it and} the network state the devices must be able to generate an arbitrary graph between all clients of the network. 

In order to distribute the network state we propose a {\it quantum network configuration protocol} (QNCP), see Sec. \ref{sec:arch:build:QNCP}. 

In a quantum network the network devices have to know which client is connected to which device, as this information is mandatory for a device to establish adjacencies via the network state, see Sec. \ref{sec:arch:build:router}. This situation can either be achieved by storing this classical information within all network devices or by storing it at specific, selected devices and querying it on demand. The latter scenario is certainly more beneficial in terms of storage. Therefore, we consider the setting where a designated router (or alternatively a classical server), has global knowledge about which client connects to which device in the network. In a classical network, routers use similar information to determine paths through networks. However, in the quantum setting which we study here, the network devices query this information from a router in order to determine which qubits of the network state need to be used for particular graph state requests. 

\begin{figure}[h!]
\begin{center}
\scalebox{3}{
\includegraphics{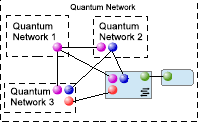}
}
\caption[h!]{\label{fig:basic:network:recursive} The quantum networks $1,2$ and $3$ connect in a larger network via quantum routers. The vertices in the figure correspond to qubits, stored within a router of each network, of the network state of the surrounding network in a GHZ architecture, see Sec. \ref{sec:arch:ghz} for details. In this example a switch connecting client $1$ is also located within the surrounding network.}
\end{center}
\end{figure}

Finally, a quantum network may also be part of a larger network. In particular, routers, see Sec. \ref{sec:arch:build:router}, may connect networks in a recursive manner, see Fig. \ref{fig:basic:network:recursive}.
For example consider the scenario of three networks, where one router resides in one network each consisting of possibly numerous switches with various clients. We refer to those networks as level $1$ networks. The routers host the network states of their respective level $1$ networks. Since routers can also connect networks, the three routers may again share a network, which we refer to as level $2$ network. This network has its own network state, just as the level $1$ networks do. Due to this fact it is also possible to connect switches to the level $2$ network without any further modification, see the switch in Fig. \ref{fig:basic:network:recursive}. From this we find that networks might appear in a nested fashion within our architectural approaches. In particular, the design we propose abstracts networks in the same fashion as network devices which enables for layered networks. 

\begin{figure}[h!]
\begin{center}
\scalebox{3.5}{
\includegraphics{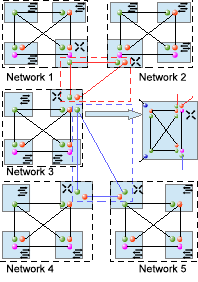}
}
\caption[h!]{\label{fig:sidebyside} Side-by-side configuration of GHZ networks, see Sec. \ref{sec:arch:ghz}. Dashed lines show networks between routers. The router of network $3$ is part of the red and blue router network, thereby he might be used as a bridge between network $1,2$ and network $4,5$.}
\end{center}
\end{figure}

It also possible to have side-by-side networks. By side-by-side networks we mean here that several quantum networks may be connected at the same hierarchical level instead of just two. This is established again by routers, which may be connected to several networks rather than one (stand-alone network) or two (hierarchical approach, see discussion above). The situation is summarized in Fig. \ref{fig:sidebyside}. 

This technique turns out to be very useful for splitting up networks with a large number of devices. In this way one can effectively reduce the size of the network state. 

Finally we highlight that only routers need to have knowledge about clients located in other networks. Switches do not need that knowledge since other networks appear at a router as {\it virtual clients}. If a client classically broadcasts a graph state request which involves clients (which are discovered classically) of other networks, the router, which connects to this network, responds to the requestor as a proxy for those clients. Therefore, the switch to which the initiator connects, establishes adjacencies to these clients via the responsible router.

We also note that due to single and two qubit measurements during target state generation the network devices need to keep track of measurement outcomes and resulting correction operations. Therefore, either the network devices ultimately apply the accumulated local correction operations after target state generation has finished or tell the clients which correction operations need to be applied to their qubits. Furthermore, we assume that classical communication is free within our architectures.

\subsection{Quantum network configuration protocols (QNCP)}\label{sec:arch:build:QNCP}

In this section we describe a network configuration protocol to distribute and extend the network state of a quantum network at an abstract level. 

Assume that a new network device starts within the network. It first checks classically if there is a QNCP server, which is responsible for the generation of the network state, is available within the network. If there is a QNCP server available and the network device connects $c_i$ clients then the client requests $c_i$ network qubits from the QNCP server. The QNCP server acknowledges this request classically with an acknowledgement message, which signals the network device that the QNCP server will now deliver the network qubits. We refer to this mode of QNCP also as server-driven QNCP. 

If no QNCP server is available, the device creates a certain state depending on the network architecture, corresponding to the possible network adjacency of his clients, and distributes parts of this state to the other network devices. Those connect the received qubits to the earlier established network state if necessary. We refer to this mode of QNCP also as device-driven QNCP. 

In Sec. \ref{sec:arch:ghz:qncp} and Sec. \ref{sec:arch:universal:qncp} we present specific protocols which implement these functionalities.

\section{Architecture A (GHZ)}\label{sec:arch:ghz}

The first architecture, which we also refer to as architecture A or GHZ architecture relies solely on GHZ states. For simplicity we do not consider any correction operations necessary due to measurement outcomes. It is straightforward to take those correction operations into account.

\subsection{Device state and protocol}\label{sec:arch:ghz:device}
Quantum clients connect to networks devices via a Bell-pair, see Fig. \ref{fig:ghz:black}. Suppose $n$ quantum clients connect to a network device. Recall that, according to Sec. \ref{sec:arch:build:switch}, the devices use the device state to generate the adjacency of a requested graph state among the directly connected clients. Therefore, the device state needs to be such that creating an arbitrary graph is feasible by manipulating this state. Hence we propose
\begin{align}
\ket{\mathrm{D}} = \bigotimes\limits^n_{i=2} \ket{\mathrm{GHZ}_{i}} \label{eq:network:ghz:local}
\end{align}
as device state for network devices. In particular, we associate the root of the GHZ state $\ket{\mathrm{GHZ}_i}$ with client $i$ for $2 \leq i \leq n$. Therefore the device needs to store $n-1$ GHZ states locally. We also refer to the root of $\ket{\mathrm{GHZ}_i}$ as the proxy qubit of client $i$. The leafs of the GHZ state $\ket{\mathrm{GHZ}_i}$ correspond to the clients $1, \ldots, i-1$. We observe that there is no root for client $1$, as the tensor product in (\ref{eq:network:ghz:local}) ranges from $2$ to $n$. Hence for client $1$ only leafs will be available, so the device chooses one leaf of the states $\ket{\mathrm{GHZ}_{i}}$ as proxy qubit for client $1$. 

\begin{figure}[h!]
\begin{center}
\scalebox{2}{
\includegraphics{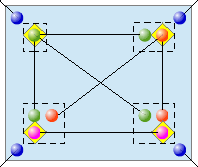}
}
\caption[h!]{\label{fig:ghz:white} The device state of a quantum network device in the setting of $4$ quantum clients and a GHZ architecture. The yellow diamonds mark the proxy qubit of the respective client. Observe that, except for client $1$, there is exactly one GHZ state for each client.}
\end{center}
\end{figure}
We observe that the network device is able to generate any $n$ qubit graph state, where $n$ corresponds to the number of clients, on the proxy qubits of the device state (\ref{eq:network:ghz:local}) as follows: First the device rotates each GHZ state according to the root and leaf configuration as specified above to the corresponding graph state (by applying Hadamard rotations), see Sec. \ref{sec:back:ghz}. Then the device applies the merging procedure of Sec. \ref{sec:back:merge:cnot} between the proxy qubit of client $i$ (root of GHZ state $\ket{\mathrm{GHZ}_{i}}$) and the leaf of the state $\ket{\mathrm{GHZ}_{j}}$ where $j \geq i$. If there is no edge in the final graph state the device measures the respective leafs of the GHZ states in the $Z$ basis. The device applies this procedure iteratively starting with client $1$ to client $n$. Finally the device teleports the proxy qubits to the clients via the respective Bell-pairs which establishes the graph state between the clients. 

Instead of sharing Bell-pairs with the clients, the device might also include the client qubits as leaf within the respective GHZ state during device state preparation, see Fig. \ref{fig:ghz:black}. This simply corresponds to the larger GHZ states $\ket{\mathrm{GHZ}_{i+1}}$ in (\ref{eq:network:ghz:local}). In that setting, the protocol for graph state generation remains the same until the very last step, since in that case the switch teleports the graph state via measurements in the $X$ or $Y$ basis to the client qubits. Observe that this reduces the storage requirement by  $n$ qubits if $n$ clients are connected. 

Because this state is a local state residing within the device, we assume that the state (\ref{eq:network:ghz:local}) may be generated within the device. Alternatively, the device may generate the device state via entanglement distillation to reach a sufficiently high fidelity of the device state relative to the state (\ref{eq:network:ghz:local}). In addition, we observe that if client $i$ is not part of the current request the state $\ket{\mathrm{GHZ}_i}$ does not need to be manipulated, i.e. it remains usable for future requests.

\subsection{Network state and protocol}\label{sec:arch:ghz:net}

As we discussed in Sec. \ref{sec:arch:build:network} we connect network devices via quantum states. Recall that the network devices use the network state to generate the adjacencies of its clients (w.r.t. the requested graph state) to clients at other network devices. In particular suppose $m$ network devices share the same quantum network. 
Furthermore assume that $c_i$ quantum clients connect to network device $i$. 

We propose the network state
\begin{align}
\ket{\mathrm{N}} = \bigotimes\limits^{m}_{i=2} \ket{\mathrm{N}_i}. \label{eq:network:ghz:nstate}
\end{align}
where
\begin{align}
\ket{\mathrm{N}_i} = \bigotimes\limits^{c_i}_{j=1} \ket{\mathrm{GHZ}_{i}}_{j}. \label{eq:network:ghz:nistate}
\end{align}
We associate network device $i$ with the state $\ket{\mathrm{N}_i}$ in (\ref{eq:network:ghz:nistate}) in order to generate the network adjacencies of clients which connect to device $i$ to clients connected to the network devices $1,\ldots,i-1$. We emphasize that several copies of GHZ states are required at each device to ensure full functionality, since multiple clients connect to each device. In particular, network device $i$ stores the $c_i$ root qubits of the GHZ states $\ket{\mathrm{GHZ}_{i}}_{j}$ of (\ref{eq:network:ghz:nistate}), one root corresponding to exactly one client of device $i$. The network devices $1, \ldots, i-1$ store one leaf qubit of each GHZ state copy in (\ref{eq:network:ghz:nistate}). We refer to the root of the state $\ket{\mathrm{GHZ}_{i}}_{j}$ also as network proxy of client $j$ at network device $i$ for the following reason: network device $i$ uses the state $\ket{\mathrm{GHZ}_{i}}_{j}$ to generate the adjacency of client $j$ at device $i$ to clients connected to the network devices $1,\ldots,i-1$. While the copies need to increase with the number of clients per device, it should be noted that the size of the GHZ states only depends on the number of devices in the network.

We roughly describe the network protocol for the GHZ architecture as follows: First the devices need to expand the GHZ states of the network ({\it state expansion phase}). After that, they connect the network GHZ states to the device GHZ states in terms of Bell-measurements ({\it state combination phase}). Finally the devices generate the required graph state by applying the connecting procedure of Sec. \ref{sec:back:merge:czconnect} and teleport the graph state to the clients ({\it state generation phase}). A detailed example of this architecture in action can be found in Appendix \ref{app:example_ghz}. Furthermore we discuss an example of a parallel request in Appendix \ref{app:example_parallel_request}.

\subsubsection{State expansion phase}

\begin{figure}[h!]
\begin{center}
\scalebox{4}{
\includegraphics{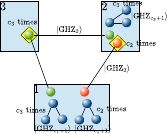}
}
\caption[h!]{\label{fig:ghz:network} Illustration of the network state for a network with three network devices, each connecting $c_1$, $c_2$ and $c_3$ clients. Observe that the devices $1$ and $2$ need to prepare the GHZ states $\ket{\mathrm{GHZ}_{c_1 + 1}}$ and $\ket{\mathrm{GHZ}_{c_2 + 1}}$ locally to connect their $c_1$ and $c_2$ clients within the network.}
\end{center}
\end{figure}

\begin{figure}[h!]
\begin{center}
\scalebox{4}{
\includegraphics{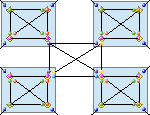}
}
\caption[h!]{\label{fig:ghz:network:example} Four devices with three clients each connected in a network. We omitted the expander states of each device for simplicity.}
\end{center}
\end{figure}

Since $c_1,\ldots,c_{i-1}$ clients connect to the network devices $1, \ldots, i-1$ respectively, the devices $1,\ldots,i-1$ have to expand their $c_i$ leaf qubits to $c_1,\ldots,c_{i-1}$ respectively. Therefore device $k$ where $1 \leq k \leq i-1$ prepares $c_i$ copies of the state $\ket{\mathrm{E}_{k}} = \ket{\mathrm{GHZ}_{c_k + 1}}$. We refer to this state also as expander state of device $k$. The situation is summarized in Fig. \ref{fig:ghz:network}.
For a concrete example of the overall state within a GHZ quantum network see Fig. \ref{fig:ghz:network:example}. 

\begin{figure}[h!]
\begin{center}
\scalebox{6}{
\includegraphics{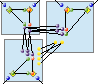}
}
\caption[h!]{\label{fig:ghz:network:localexpand} The state after expanding the network GHZ states.}
\end{center}
\end{figure}

The state expansion phase for establishing a fully connected graph state within the network (the protocol for establishing arbitrary graph states is just a slight modification of it) is now as follows: Each device $1 \leq k \leq m$ connects its expander states $\ket{\mathrm{E}_{k}}$ to its leafs of the GHZ states $\ket{\mathrm{GHZ}_{l}}^{\otimes c_l}$ where $k+1 \leq l \leq m$ of the network states via a Bell-measurements. This results in a tensor product of expanded GHZ states, see Sec. \ref{sec:back:ghz}. More specifically, after the expansion step the parts of the network states are given by
\begin{align}
\ket{\mathrm{N}'_i} = \bigotimes\limits^{c_i}_{j=1} \ket{\mathrm{GHZ}_{1+\sum^{i-1}_{k=1} c_k}}_{j} \label{eq:network:ghz:nistatenew}
\end{align}
where $2 \leq i \leq m$. The states $\ket{\mathrm{GHZ}_{1+\sum^{i-1}_{k=1} c_k }}_{j}$ of (\ref{eq:network:ghz:nistatenew}) now enable to generate adjacencies between the clients of device $i$ and clients connected to devices $1,\ldots,i-1$, see Fig. \ref{fig:ghz:network:localexpand}. 

We observe that $\ket{\mathrm{GHZ}_{1+\sum^{i-1}_{k=1} c_k }}_{j}$ within (\ref{eq:network:ghz:nistatenew}) enables the generation of edges between client $j$ of device $i$ to clients which connect to the devices $1,\ldots,i-1$. More precisely, by recalling that the leafs of the state $\ket{\mathrm{GHZ}_{1+\sum^{i-1}_{k=1} c_k }}_{j}$ stem from Bell-measurements within the devices $1,\ldots,i-1$ we find that these devices may use their $c_k$ leafs to generate an edge between client $j$ at device $i$ and the clients $1 \leq l \leq c_k$ at device $k$. Consequently, if no edge needs to be established, all devices $k$ measures this leaf qubit in the $Z$ basis. 

Alternatively, and certainly more beneficial in terms of required storage capacity, device $k$ where $1 \leq k \leq i-1$ could prepare smaller expander states $\ket{\mathrm{E}_{k}}$. For example, if the part of the network state $\ket{\mathrm{GHZ}_{l}}$ corresponds to the adjacency of client $i$ at device $l$, and if only two clients of device $l-1$ are adjacent to client $i$ in the final graph, it suffices for device $l-1$ to prepare the expander state $\ket{\mathrm{GHZ}_3}$ instead of $\ket{\mathrm{GHZ}_{c_{l-1}+1}}$.

\subsubsection{State combination phase}

\begin{figure}[h!]
\begin{center}
\scalebox{4}{
\includegraphics{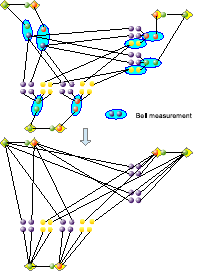}
}
\caption[h!]{\label{fig:ghz:network:networkmerge} In this example the right switch, i.e. switch $2$, creates two $\ket{\mathrm{GHZ}_{3}}$ as adapter states, one for each client, whereas the lower switch creates two $\ket{\mathrm{GHZ}_{4}}$ states as adapter states. Then the GHZ network states get connected to the local GHZ states via Bell-measurements (indicated by ellipses). Finally each switch generates the adjacency of the requested graph state by employing connect procedures, see Sec. \ref{sec:back:merge:czconnect}, i.e. controlled phase gates and $Y$ measurements.}
\end{center}
\end{figure}

Recall that $c_i$ clients connect to network device $i$. The device state of device $i$, i.e. $\ket{\mathrm{D}_i}$, is given by (\ref{eq:network:ghz:local}). To account for adjacencies of the $c_i$ clients across the network we extend the device state $\ket{\mathrm{D}_i}$ by the virtual client $1$, corresponding to the network. Observe that according to Sec. \ref{sec:arch:ghz:device} this virtual client only has leaf qubits of the device state. This results in the new device state
\begin{align}
\ket{\mathrm{D}_i} = \bigotimes^{c_i + 1}_{j=2} \ket{\mathrm{GHZ}_{j}}. \label{eq:ghz:newdevice}
\end{align}
The protocol proceeds as follows: Since the expanded network state $\ket{\mathrm{N}'_i}$ of (\ref{eq:network:ghz:nistatenew}) needs to be finally connected with the device state $\ket{\mathrm{D}_i}$ of (\ref{eq:ghz:newdevice}), each device $i \neq m$ creates $c_i$ copies of the GHZ state $\ket{\mathrm{GHZ}_{2+\sum^{m}_{k=i+1} c_k}}$, one for each of its client, as adapter for the device GHZ states. Observe that exactly $1 + \sum^{m}_{k=i+1} c_k$ leafs are necessary, as there are $\sum^{m}_{k=i+1} c_k$ leafs from the network states $\ket{\mathrm{N}'_k}$ for $i+1 \leq k \leq m$ stored at device $i$ and the roots of the network GHZ state $\ket{\mathrm{N}'_i}$. First the devices connect the root of the states $\ket{\mathrm{N}'_i}$, i.e. the network proxies of the clients, to the adapter states, thereby expanding the adapter state. Then each device performs a Bell-measurement between the $c_i$ virtual network leafs of the device state (\ref{eq:ghz:newdevice}) and the expanded adapater states, resulting in an expanded device state, see Fig. \ref{fig:ghz:network:networkmerge}. This procedure enables device $i$ to generate edges to the clients of device $i+1, \ldots, m$. 

Also in this case, similar to the state expansion phase of the previous section, the device could generate smaller adapter states in principle. In particular, the leafs of the adapter state $\ket{\mathrm{GHZ}_{2 + \sum^{m}_{k=i+1} c_k}}$, see Fig. \ref{fig:ghz:network:networkmerge}, will ultimately be used to generate the network adjacency of the clients of a device. Therefore, depending on the graph state request, the device can prepare smaller adapter states tailored to the network adjacency of the device. For example, if a client of device $i$ is adjacent to two clients on devices $i+1, \ldots, m$ and to clients at devices $1, \ldots, i-1$, then it suffices to prepare the adapter state $\ket{\mathrm{GHZ}_{4}}$ instead of $\ket{\mathrm{GHZ}_{2 + \sum^{m}_{k=i+1} c_k}}$.


\subsubsection{State generation phase}

In this final phase, the devices rotate the GHZ states of the previous phase according to their root and leaf configuration to graph states via Hadamard rotations, see Sec. \ref{sec:back:ghz}. Now the devices generate the requested graph state by performing connecting procedures according to Sec. \ref{sec:back:merge:czconnect} between the leafs of the device states and the leafs of the network GHZ states. The adjacency of clients connected to a device is generated according to the protocol in Sec. \ref{sec:arch:ghz:device}. This generates the graph state on the proxy qubits of the clients. 

Finally the network devices teleport the proxy qubits to the clients via the Bell-pairs to the clients. 

\subsubsection{Networking}
Recall that routers may connect several networks. In particular, a router is a switch with the ability to connect networks, and to generate and distribute the state of the network $\ket{\mathrm{N}}$ of (\ref{eq:network:ghz:nstate}). 

\begin{figure}[h!]
\begin{center}
\scalebox{4}{
\includegraphics{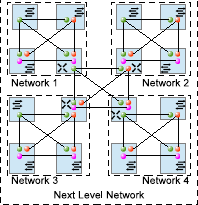}
}
\caption[h!]{\label{fig:ghz:network:multi} The figure depicts the GHZ architecture applied recursively. The routers connect the networks $1,2,3$ and $4$ within a network on the next level, again, using GHZ network states.}
\end{center}
\end{figure}

Therefore it is straightforward to extend the previous network protocol to a setting where quantum networks connect via routers: By employing the same architecture, i.e. the same network state and network protocol, as described above between several routers the routers share a network with network state (\ref{eq:network:ghz:nstate}). We only need more GHZ states in (\ref{eq:network:ghz:nistate}) to account for all clients of the network. This implies that routers which connect two or more networks enable the generation of graph states between clients located within different networks, see Fig. \ref{fig:ghz:network:multi}. Also observe that this approach enables recursive networks. 

In particular, router $R$ abstracts network $N$ w.r.t. to other networks as follows: clients located within $N$ can reach other networks via router $R$. Clients outside of network $N$ can reach clients located within network $N$ only via router $R$. Therefore, to other networks, network $N$ appears simply as a network device connecting all clients of network $N$. Hence the same states as in (\ref{eq:network:ghz:nistate}) need to be distributed among the routers, just by replacing the number of copies of GHZ states with all clients located within a particular network, i.e. $\sum^m_{i=1} c_i$. 


We also want to point out that, identical to device states, if certain devices, or more precisely all the clients of a device, are not involved in a request, the corresponding GHZ states of the network can be kept by the devices. Therefore, depending on the request, it may not be necessary to distribute the whole network state for every request in the network.

\subsection{Security considerations}\label{sec:arch:ghz:sec}

We reiterate that within the GHZ network protocol an edge appears in the generated graph state if a connecting procedure, see Sec. \ref{sec:back:merge:czconnect}, is done. In order to remove an edge (or not to establish an edge) the devices have to measure the leaf qubits of the extended network GHZ state (\ref{eq:network:ghz:nistatenew}) in the $Z$ basis. Observe that in both cases each device needs to operate on the network state. Hence we need to distinguish two different security settings: either we assume that all network devices are trustworthy, i.e. they execute the protocol, or at least one devices does not. 

Before providing a discussion about the security of our architectures we first recall the definition of security for entanglement distillation protocols given in \cite{Pirker16a, Zwerger17}. There, the security of an entanglement distillation protocol is defined as the distance of the output states between a real distillation protocol (for a finite number of initial states) and an ideal distillation protocol maximized over all purifications of initial states (where the purifications are held by an eavesdropper). The real distillation protocol consumes a finite number of initial states outputs a mixed state with fidelity  close to unity. The output state of the ideal distillation protocol is, in case of a noiseless distillation protocol, one or several copies of a pure states (for bipartite distillation protocols one or several Bell-pairs, for multipartite distillation protocols one or several copies of a graph state). The results of \cite{Zwerger17} for the hashing protocol imply that the security of the hashing protocol is bounded by the trace distance between the output of the hashing protocol and several copies of the pure target state, which translates to the fidelity of the output of the hashing protocol. Furthermore it was shown that the security parameter scales exponentially fast towards zero in terms of the number of initial states, thereby implying that the hashing protocol guarantees exponential security levels. The protocol of \cite{Zwerger17} is discussed in more detail in section \ref{sec:sec:trusted}.

\subsubsection{Security considerations -- trusted networks}\label{sec:sec:trusted}

First we treat the case of a trusted network. Observe that noise due to transmission decrease the fidelity of network states relative to pure GHZ states. In order to deal with this channel noise we employ entanglement distillation protocols for two colorable graph states to generate high fidelity network states \cite{Du05,EPPallGraphs,Duer07,Miyake05,Fujii09,Chen07,ManevaEPP,HostensEPPGraph,GlancyEPPGraph,HostensEPPGraph2}. Recall that the requested graph state is generated by connecting GHZ states (which are two-colorable) via the connecting procedures of Sec. \ref{sec:back:merge:czconnect} which the network devices apply. 

Using entanglement distillation to create high fidelity states has an advantage for implying security \cite{Pirker16a, Zwerger17}. In \cite{Zwerger17} the security of a noisy measurement-based implementation of multipartite hashing protocols was investigated in scenarios where an eavesdropper distributes all initial states subject to distillation. Hashing protocols operate on a large noisy ensemble of initial states in a collective manner, where parity information about the noisy ensemble is being learnt, thereby purifying it. It was shown that if multipartite hashing protocols are prepended by symmetrization, a twirl (towards diagonal states w.r.t. to the graph state basis) and a parameter estimation, then hashing factors out the eavesdropper even in the presence of noise and imperfections. In particular, these modified hashing protocols converge towards states which can be described by local depolarizing noise acting on several copies of pure graph states, where noise stems solely from noise in the apparatus. This implies that the final states after hashing remain private, i.e. the hashing protocol disentangles any eavesdropper.  

We use this observation to imply the security of our protocol in the case where all network devices are trustworthy as follows: All network devices receive their network GHZ states from a QNCP server (or establish them via device-driven QNCP). The device state residing within every network device is generated by the devices themselves, or distilled from initial states of an eavesdropper. In the former case security is guaranteed since the devices are trustworthy, whereas in the latter case the devices establish local security at a device level via entanglement distillation. In addition, we employ entanglement distillation after routing (which corresponds to the process of measuring a leaf qubit in the $Z$ basis if it is not part of the request) of the network states. Hence, in the trustworthy scenario we can assume that all not involved network devices measure their leaf qubits of the GHZ network states in the $Z$ basis. The devices which are involved in a request then run entanglement distillation on the routed network states, thereby destroying any entanglement with a potential eavesdropper outside the network. This process is performed for every graph state request. Since the security definition of \cite{Zwerger17} is composable, the security of the network states and the security of the device states imply security at the device and network level. 

Finally the devices teleport the generated graph state to the clients. The Bell-pair used for this teleportation process also stems from entanglement distillation, and, as shown in \cite{Pirker16a}, bipartite entanglement distillation protocols, even when carried out via imperfect quantum gates, lead to security. Therefore the Bell-pair can be assumed to be private, which finally implies via composability that the established graph state is secure. However, this also means that the clients need to possess the additional capability to perform a bipartite entanglement distillation protocol, if this connection to their nearest quantum network device is untrusted. This requires two qubit entangling gates at the clients.

\subsubsection{Security considerations -- untrusted networks}

\begin{figure}[h!]
\begin{center}
\scalebox{4}{
\includegraphics{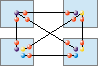}
}
\caption[h!]{\label{fig:ghz:network:enhanced} The figure illustrates the secure GHZ network protocol state. We decorate each edge of the network GHZ states with two additional qubtis (red vertex), where one of each resides within the device. This enables the device to either establish an edge ($Y$ measurement) or to remove an edge ($Z$ measurement) locally. This procedure removes the trust relationship to other, not involved, network devices.}
\end{center}
\end{figure}

In cases where the network itself is not trustworthy, we modify the network states as follows: we decorate each edge of the GHZ states of (\ref{eq:network:ghz:nistatenew}) with two additional qubits. One of this decorators belongs to the root of the GHZ state, the other one to the leaf of the GHZ state, see Fig. \ref{fig:ghz:network:enhanced}.
This enables the devices to measure the decorator qubits of needed edges of the network states in the $Y$ basis (which establishes an edge from the root of the GHZ state to the leaf) and to measure all unneeded edges by a measurement in the $Z$ basis, which removes this edge. This has the advantage that the network devices of a request do not have to trust other, not involved, network devices. We note that the procedure above results in GHZ states, which enable to further proceed with the GHZ network protocol of Sec. \ref{sec:arch:ghz:net}. This implies security at the device and network level via the same arguments as above, i.e. entanglement distillation. We refer to this modified protocol also as secure GHZ network protocol. 

In situations where all network devices are untrusted one implies the security claim of the protocol via the following arguments. Recall that there exist entanglement distillation protocols for all graph states \cite{EPPallGraphs}. Further assume that all clients which are part of the graph state request demand the network to receive several qubits from their respective network devices, assuming that each qubit is part of a copy of the target state. By extending the computational capabilities of the clients to run the distillation protocol for all graph states of \cite{EPPallGraphs} they may generate privacy via entanglement distillation. Therefore, by appending the entanglement distillation protocol of \cite{EPPallGraphs} after the network protocol is finished, one establishes security even if all network devices are untrustworthy. Such a functionality corresponds to classical firewalls, which protect clients from network attacks. 

On the one hand this approach has the advantage that no assumptions on the network devices is necessary, but on the other hand, the clients need to be capable of running the entanglement distillation protocol of \cite{EPPallGraphs}. In addition they have to store several qubits at the same time, as those will be subject to distillation. This observation implies a memory overhead for each client. Notice that these additional capabilities of the client are only necessary if the whole network or their connection to their nearest quantum network device is untrusted.

\section{Architecture B (decorated-type)}
\label{sec:arch:universal}

The second architecture we propose relies on a different class of states: decorated graph states, which we introduce briefly. Also in this section we do not consider any correction operations necessary due to measurement outcomes for simplicity.

\subsection{Device state and protocol}\label{sec:arch:universal:device}

We recall our setting as discussed in Sec. \ref{sec:arch}: quantum clients connect to switches and/or routers via quantum channels from which they receive a qubit which will finally be part of the request graph state. 

\begin{figure}[h!]
\begin{center}
\scalebox{4}{
\includegraphics{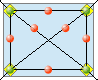}
}
\caption[h!]{\label{fig:universal:local} The figure depicts the device state of a network device following the decorated architecture. The green vertex correspond to the proxy qubits whereas the red vertex to the decorator qubits.}
\end{center}
\end{figure}

The device state within a decorated architecture can be described as follows: the network device stores one qubit for each client (which we refer to also as proxy qubit, see also proxy qubits within the GHZ device state in Sec. \ref{sec:arch:ghz:device}). The network device attachs to this proxy qubit another qubit which is send to the client. The client proxies are fully connected among each other and we decorate each edge within this graph state by a qubit (which we also refer to as decorator qubit). The device uses the decorator qubit to either establish or remove an edge within the final graph state, see Fig. \ref{fig:universal:local}, via a measurement in the $Y$ basis or the $Z$ basis respectively. 

The protocol to generate an arbitrary graph state between the clients is now as follows: The network creates the graph state via measurements in the $Y$ and $Z$ basis. Then the device teleports the graph state via $X$ and $Y$ measurements of the proxy qubits to the clients.

\subsection{Network state and protocol}\label{sec:arch:universal:net}

Here we discuss the network state and protocol for untrusted networks, since for trusted networks the protocol remains the same and its network state is easily obtained from the network state of untrusted networks. 

\begin{figure}[h!]
\begin{center}
\scalebox{4}{
\includegraphics{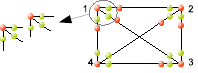}
}
\caption[h!]{\label{fig:uni:network} The decorated network state for a network of four devices. Observe that the resulting state is two colorable.}
\end{center}
\end{figure}

We connect the network devices via a similar state as in Fig. \ref{fig:universal:local}. Suppose $m$ devices connect within a quantum network. Instead of sharing a fully connected graph state, the $m$ devices connect via an $m$ partite graph where each partition $i$ contains $c_i$ qubits, one for each client. We refer to these qubits of the network state, similar as in Sec. \ref{sec:arch:ghz:net}, as network proxies, since they will finally be used to generate the network adjacencies of the clients. Furthermore we decorate each edge within this $m$ partite graph as follows: network device $i$ decorates the edges of its partition to the partition of the network devices $1,\ldots,i-1$ once and edges to network devices $i+1,\ldots,m$ twice, see Fig. \ref{fig:uni:network}. This decoration procedure ensures that the resulting state is a two-colorable graph and, more importantly, that each network device can either establish or remove edges within the network state without assuming a particular action of the neighbouring network devices. We will discuss this in more depth at the end of this subsection. 

\begin{figure}[h!]
\begin{center}
\scalebox{4}{
\includegraphics{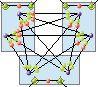}
}
\caption[h!]{\label{fig:uni:example:1}The figure illustrates an example of a decorated network. The device states and the network state both rely on  decorated states. In this example we consider a network of three devices where each connects two clients.}
\end{center}
\end{figure}

Within this architecture the generation of an arbitrary graph state is achieved by combining the device protocol, see Sec. \ref{sec:arch:universal:device}, and the network protocol we discuss below. An example network is shown in Fig. \ref{fig:uni:example:1}. 

\begin{figure}[h!]
\begin{center}
\scalebox{4}{
\includegraphics{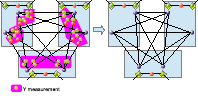}
}
\caption[h!]{\label{fig:uni:step1}The first step is to measure all decorator qubits of edges which contribute to the requested graph state in the $Y$ basis. We measure the decorator qubits of unwanted edges in the $Z$ basis. In this example a fully connected graph state needs to be established.}
\end{center}
\end{figure}

\begin{figure}[h!]
\begin{center}
\scalebox{4}{
\includegraphics{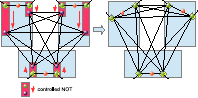}
}
\caption[h!]{\label{fig:uni:step2} The devices apply the merging procedure, i.e. a CNOT with the proxy qubits as source and the network qubits as target followed by measurement in the $Z$ basis of the network qubits. This transfers the network adjacency from the network qubits to the proxy qubits.}
\end{center}
\end{figure}

The network devices establish the network adjacency of their clients on the network proxies by either measuring their decorator qubits in the $Y$ basis, which establishes an edge, or in the $Z$ basis, which removes an edge, see Fig. \ref{fig:uni:step1}.
Recall that the device state contains one proxy qubit for each client. In addition, recall that the network state also contains exactly one network proxy qubit for each client. The devices now apply to each client-wise pair of device proxy and network proxy qubit the merging procedure of Sec. \ref{sec:back:merge:cnot} with the device proxy as source and the network proxy as target. The steps are summarized in Fig. \ref{fig:uni:step2}. After measuring the decorator qubits of the device states the device proxy qubits form the requested graph state which enables the network device to transfer the graph state to the clients by measuring the device proxies accordingly. 

Observe that also this approach may again be easily applied recursively, similarly as for the GHZ architecture in Sec. \ref{sec:arch:ghz:net}. Identical as to the GHZ architecture one abstracts whole networks as black-boxes which appear as a router at the next level. The network state at the next level is again a $k$ partite graph state, but now, the qubits within that partition need to be equal to the number of clients of the abstracted network. 


One possibility to deal with large network states may be to split up large networks into smaller networks, see Sec. \ref{sec:arch:build:network}, as this leads to smaller network states. We observe that, as we proposed the architecture of this section, only a single copy of the decorated network state is present in the network. Therefore another possibility to deal with large networks is to prepare several copies of smaller decorated states which may be merged upon request by the network devices.

Recall that the network devices share a decorated network state, see Fig. \ref{fig:uni:network}. Suppose a request is broadcasted within the network. Then the network devices which are part of that request measure the decorator qubits of unwanted edges in the $Z$ basis. Thereby they remove correlations to other network devices which are not part of the request, see Sec. \ref{sec:arch:ghz:sec} for untrusted networks. Observe that the resulting state still remains two colorable.

Now security for this architecture in untrusted networks follows via the same arguments as in Sec. \ref{sec:arch:ghz:sec}: The network state after routing, i.e. removing unwanted edges via measurements in the $Z$ basis, remains two colorable and we assume that the network state is created in terms of entanglement distillation. Therefore the security of this architecture follows from the discussion of Sec. \ref{sec:arch:ghz:sec}.

Note that a simpler network state with only one decorating qubit per edge can be used if the ability to securely disconnect non-participating devices is not needed. This might be the case for trustworthy networks where all network devices act according to the network protocol described in this subsection.

\section{Architecture C (hybrid)}\label{sec:arch:hybrid}

So far we had a sharp distinction between GHZ networks and networks based on fully connected decorated graph states. Both architectures differ in the kind of states they use, see Fig. \ref{fig:ghz:white} and Fig. \ref{fig:universal:local} for the device state and Fig. \ref{fig:ghz:network} and Fig. \ref{fig:uni:network} for the network state. 

\begin{figure}[h!]
\begin{center}
\scalebox{6}{
\includegraphics{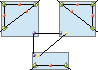}
}
\caption[h!]{\label{fig:hybrid:example} Graphical illustration of a hybrid architecture: the devices have the device state as discussed in Sec. \ref{sec:arch:universal:device} whereas the devices within the network connect via the network GHZ states (\ref{eq:network:ghz:nistate}).}
\end{center}
\end{figure}

An example of a hybrid network is depicted in Fig. \ref{fig:hybrid:example}. 

Observe that also in this case the generation of arbitrary graph states is possible. One just needs to slightly modify the protocol of Sec. \ref{sec:arch:ghz:net} or Sec. \ref{sec:arch:universal:net}. Instead of the final Bell-measurement between network proxy qubits and network qubits one applies a CNOT as described in Sec. \ref{sec:arch:universal:net}. Similarly one easily derives a protocol for GHZ device states and decorated network states. 

Therefore we find that any combination of device and network states might appear within our architectures. This implies that our solutions offer a high degree of modularity, since the network may be composed of different device architectures at the same time.

\section{Architecture D (Bell-pairs)}\label{sec:arch:bellpairs}
 
We also discuss how one may establish the same functionality as we offer via the architectures of Sec. \ref{sec:arch:ghz} and Sec. \ref{sec:arch:universal} directly via Bell-pairs. In particular, we assume the same building blocks as proposed in Sec. \ref{sec:arch}, i.e. clients which solely connect to switches or routers. Furthermore we follow the same top-down approach: the network state, which now consists only of Bell-pairs between the network devices, is distributed prior to any graph state request. 

\begin{figure}[h!]
\begin{center}
\scalebox{4}{
\includegraphics{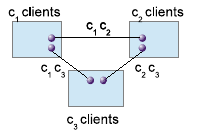}
}
\caption[h!]{\label{fig:bipartite:network} The figure shows the setting when three network devices connect via Bell-pairs. The devices connect $c_1$, $c_2$ and $c_3$ clients respectively. Therefore $c_1 c_2$, $c_1 c_3$ and $c_2 c_3$ Bell-pairs need to be distributed between the respective devices.}
\end{center}
\end{figure}

When one uses Bell-pairs instead of multipartite states to generate the network state then, if $c_i$ clients to device $i$ and $c_{i+1}$ clients connect to device $i+1$, one needs $c_{i} c_{i+1}$ Bell-pairs between those devices to generate arbitrary edges between clients located at device $i$ and $i+1$. It can not be less since each client of device $i$ may have an edge to a client located at device $i+1$ which needs to be dynamically established. Hence to generate the adjacency of a single client of device $i$ we need at least $c_{i+1}$ Bell-pairs. In total we therefore need $2 c_{i} c_{i+1}$ qubits to establish arbitrary edges between clients which connect to device $i$ and $i+1$, see Fig. \ref{fig:bipartite:network}. As we discuss in Sec. \ref{sec:compare} this introduces an overhead of almost $2$ compared to the architectures in Sec. \ref{sec:arch:ghz} and Sec. \ref{sec:arch:universal}. 

Nevertheless, the protocol for establishing arbitrary graph states via Bell-pairs is similar to the protocol of Sec. \ref{sec:arch:ghz}: the network devices establish edges across devices by first expanding their device state Bell-pairs with a GHZ state as an adapter state. Then, depending on the adjacency of the request, they either couple a Bell-pair to these GHZ adapter states or not. Finally, the merging procedures of Sec. \ref{sec:back:merge} need to be applied to establish the adjacency across the network. 

We point out that there might also exist more optimized schemes relying solely on Bell-pairs. For example, one strategy could be to generate a Bell-pair between network devices where one half of the Bell-pair is then coupled to a locally created GHZ state in terms of a Bell-measurement. Observe that this results again in a GHZ state as it simply teleports one qubit of the GHZ to the remote network device. 

A further optimization corresponds to having a single master node as in \cite{Cuquet2012}. This reduces the number of Bell-pairs to $c$ if $c$ clients are within a given network. In such a scenario, the single master node needs to generate the graph states of given requests which the node then teleports to the clients. 

\section{Quantum network configuration protocols}\label{sec:qncp}

In this section we discuss the quantum network configuration protocols for architecture A (GHZ) and architecture B (decorated). Recall that these protocols are responsible for distributing and extending the network state of the respective architectures, see Sec. \ref{sec:arch:build:QNCP}.

\subsection{QNCP for GHZ architecture}\label{sec:arch:ghz:qncp}

\begin{figure}[h!]
\begin{center}
\scalebox{2.5}{
\includegraphics{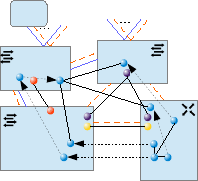}
}
\caption[h!]{\label{fig:basic:routernet} The QNCP server delivers the network qubits via Bell-pairs in terms of teleportation.}
\end{center}
\end{figure}

First we discuss the server-driven QNCP for GHZ architectures. The client checks whether a QNCP server is available or not, see Sec. \ref{sec:arch:build:QNCP}. If yes and there are $m$ devices within the network, the QNCP server answers the configuration request by creating $c$ copies of the GHZ state $\ket{\mathrm{GHZ}_{m + 1}}$, where $c$ corresponds to the number of clients connected to the new device. Next the QNCP server sends the roots of the GHZ states to the new device and the leafs to the other network devices using the available quantum channels, see Fig. \ref{fig:basic:routernet}.
This approach has the advantage that the previously distributed network state will not be destroyed if new network devices appear within the network, i.e. they can still be used for graph state generation. 

Now we discuss the device-driven QNCP for GHZ networks. In this case, the start-up procedure for the network devices is slightly different than within the server-driven QNCP protocol. Any new network device creates $c$ copies of the GHZ state $\ket{\mathrm{GHZ}_{a+1}}$ on boot locally, i.e. a GHZ state with $a$ leafs where $a$ corresponds to the number of neighbours of the new device. Then, the network device sends the $a$ leafs of the $c$ GHZ states over the quantum channels to its neighbour devices. This establishes $c$ copies of GHZ states within the neighbourhood of the new network device. 

Next the devices which receive the leafs check all their respective neighbours if they also received leafs during the previous round. If all neighbours have received leafs, the procedure completes for this network device. Otherwise, if $a'$ neighbours do not have received leafs, then the network device creates $c$ copies of a GHZ states with $a'+1$ leafs, i.e. the state $\ket{\mathrm{GHZ}_{a'+2}}^{\otimes c}$, connects the roots of these GHZ states to earlier received leafs (thereby enlarging the GHZ state of the first step), and keeps one leaf each locally. The device broadcasts the remaining leafs to the $a'$ so far unconnected neighbours. 

\begin{figure}[h!]
\begin{center}
\scalebox{4}{
\includegraphics{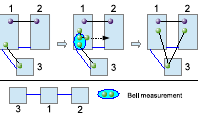}
}
\caption[h!]{\label{fig:basic:cQNCP} Illustration of device-driven QNCP: In this example we consider three network devices, $1,2$ and $3$ respectively, where devices $1$ and $2$ as well as $1$ and $3$ share a quantum channel. Suppose device $1$ and $2$ boot first. According to device-driven QNCP they connect via a Bell-pair ($2$ qubit GHZ state). Next device $3$ starts. It directly connects to device $1$ via a Bell-pair. In order to establish a connection to device $2$, device $1$ creates a three qubit GHZ state. One of the leafs is kept locally at device $1$, the second leaf is send via the quantum channel to device $2$. Finally, device $1$ connects the root of the GHZ state with the received qubit of the Bell-pair, thereby enlarging the GHZ state over the whole network.}
\end{center}
\end{figure}

The clients within the network repeat this procedure recursively until each client receives $c$ leafs of the new GHZ network state. We note that this procedure ends after at most $m$ recursions where $m$ corresponds to the number of network devices within the network. An example of the device-driven QNCP is depicted in Fig. \ref{fig:basic:cQNCP} for three network devices. 

\begin{figure}[h!]
\begin{center}
\scalebox{2}{
\includegraphics{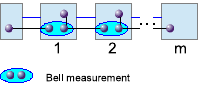}
}
\caption[h!]{\label{fig:basic:cQNCP:linear} Illustration of the device-driven QNCP protocol within a linear chain channel configuration.}
\end{center}
\end{figure}

This approach for establishing the network state inherently consumes resources. The ideal channel configuration is if all devices within the network are fully connected via quantum channels. In contrast, the worst case within that setting corresponds to a linear chain. More precisely, if a new device connects to one of  the outermost device in a linear chain of $m$ devices, then each formerly device within the chain needs to create a three qubit GHZ state. The new device connects to its neighbouring device via a Bell-pair. In total there are $m-1$ three qubit GHZ states and a Bell-pair necessary to establish the new network state, which amounts to $3(m-1) + 2$ qubits and $m-2$ Bell-measurements. For a graphical illustration see Fig. \ref{fig:basic:cQNCP:linear}. 

We also want to emphasize that the distribution of the network state over long distances could also rely on quantum repeaters, and high-fidelity generation in noisy settings can be achieved using entanglement distillation.

\subsection{QNCP for decorated architecture}\label{sec:arch:universal:qncp}

Now we present a protocol for distributing and extending the network state within decorated architectures. 

Suppose the network state within a decorated network has been established. In particular, the state depicted in Fig. \ref{fig:uni:network} is distributed in the network. Further assume, that in total there are $m$ network devices where network device $i$ connects $c_i$ clients. 

If now a new device, i.e. device $m+1$, connects to the network we need to modify the network state accordingly to take the new network device and its clients into account. For simplicity we illustrate the protocol for one client connected to device $m+1$, as the protocol just needs to be repeated to account for an arbitrary number of clients.

We propose the following protocol to modify the network state without destroying it: If the new device connects one clients, then, either the QNCP server or the new network device, creates one decorated GHZ states of size $1 + \sum^m_{j=1} c_j$. It is the same as the decorated GHZ states shown in Fig. \ref{fig:ghz:network:enhanced} but with only one decorating qubit instead of two. This state is directly associated with the client of network device $m+1$ whereas the leafs of the decorated GHZ state correspond to the clients located at the network devices $1, \ldots, m$. 

\begin{figure}[h!]
\begin{center}
\scalebox{4}{
\includegraphics{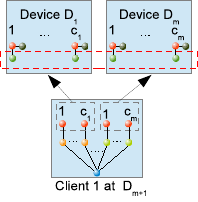}
}
\caption[h!]{\label{fig:universal:qncp} QNCP for decorated networks: If a new device starts connecting one client, then a decorated GHZ state of size $1 + \sum^m_{j=1} c_j$ needs to be prepared. In order to connect the decorated GHZ state to the current network state the leafs of the decorated GHZ state will be send to the corresponding network devices. The devices create a decorated edge on the previous network state (red dashed line) and connect the incoming qubits corresponding to the new device via CZ gates.}
\end{center}
\end{figure}

Next, each network device $1,\ldots,m$ modifies his part of the network state as follows: it creates locally for each client proxy one qubit in the $\ket{+}$ state. Then he connects these qubits to the proxy qubits via CZ gates, thereby establishing a new edge within the network state. Next the leafs of the decorated GHZ states will be send to corresponding network devices. The situation is depicted in Fig. \ref{fig:universal:qncp}. 

Finally the devices $1, \ldots, m$ connect their received leaf qubits, corresponding to connections to the new network device $m+1$ to their previously generated decorator qubits via CZ gates.

\section{Comparison of the architectures}\label{sec:compare}

In order to determine the efficiency of the different architectures we compare the multipartite approaches A and B to architecture D which relies solely on Bell-pairs. Schemes relying only on bipartite entangled states are a common scenario considered for quantum repeater networks \cite{Meter2013a,Meter2013b,Pant17,Pant17a}. In particular we determine the performance of our architectures by computing two parameters: number of qubits to be stored and the maximum number of required operations to establish an arbitrary graph state. In our approach, we can freely decide for each edge in the graph if it should appear or not. The worst case in such an approach is given by the fully connected graph state (if created in a naive manner), since all edges need to be generated. That is, the number of gates and measurements which have to be applied is at most. Notice that the fully connected graph state is LU equivalent to a GHZ state, which could be generated more efficiently.

\subsection{Device level}

We compute the number of qubits which the device needs to store in architecture A as follows: According to (\ref{eq:network:ghz:local}) the device state is given by $\bigotimes^n_{i=2} \ket{\mathrm{GHZ}_{i}}$. The number of qubits which the network device needs to store is thus given by
\begin{align}
\sum\limits^n_{i=2} i = \sum\limits^n_{i=1} i - 1 = n(n+1)/2 - 1.
\end{align}
For the number of local operations to establish a fully connected graph state, we observe the following: Recall that the root of the GHZ state $\ket{\mathrm{GHZ}_{i}}$ is the proxy qubit of client $i$, which the device teleports to the client after completing the graph state generation on the proxy qubits. According to the protocol description the device generates the adjacency by employing the merging procedure of Sec. \ref{sec:back:merge:cnot}, i.e. CNOT gates between leafs and roots of GHZ states followed by $Z$ measurements. Therefore, the number of CNOT gates is equal to the number of leafs of the GHZ states of (\ref{eq:network:ghz:local}) minus $1$ (since the leaf of the GHZ state $\ket{\mathrm{GHZ}_{2}}$ can be used as proxy of client $1$), i.e.
\begin{align}
\sum\limits^{n-1}_{i=2} i = \sum\limits^{n-1}_{i=1} i - 1  = \frac{n(n-1)}{2} - 1. \label{eq:network:ghz:locc}
\end{align}
After applying the CNOT gates the device measures the leafs of the GHZ states of (\ref{eq:network:ghz:local}) in the $Z$ basis which implies for the total number of measurements the expression of (\ref{eq:network:ghz:locc}). 

The number of qubits which need to be stored for architecture B is determined as follows: recall that the device state corresponds to a fully connected, decorated graph state. Therefore we the number of qubits corresponds to the number of clients, i.e. $n$, plus the number of edges of a fully connected graph, i.e. $n(n-1)/2$. This also corresponds to the number of measurements in the $Y$ basis in order to generate the fully connected graph among the clients.

Using Bell-pairs the network device needs to store one Bell-pair for each possible edge within the graph state. Since the number of edges within the fully connected graph state for $n$ particpants is $n(n-1)/2$ the network device needs to store $n(n-1)$ qubits. The number of controlled NOT gates is determined as follows: Observe that within the Bell-pair approach each client proxy has exactly $n-1$ qubits, because each client has $n-1$ possible neighbours. One of these qubits will be used within the final teleportation process, hence there are $n-2$ qubits left which need to be connected to the proxy before teleportation. This implies for the total number of merging procedures $n(n-2)$.  

\begin{table}[h!]
\begin{tabular}{l|c|c|c}
& D & A & B  \\
\hline Qubits & $n(n-1)$ & $n(n+1)/2 - 1$ & $n(n+1)/2$ \\
CNOT & $n(n-2)$ & $n(n-1)/2 - 1$ & - \\
Meas. & $n(n-2)$ & $n(n-1)/2 - 1$ & $n(n-1)/2$
\end{tabular}
\caption{\label{tab:local} The table summarizes the number of qubits which need to be stored within each architecture and the number of operations to establish the fully connected graph state at the device level.}
\end{table}
We summarize our findings on the device level within Table \ref{tab:local}. We find that the number of qubits which have to be stored within the proposed architectures is approximately half as in the direct approach using Bell-pairs for large $n$, see also Table \ref{tab:local:values}. Furthermore, the same holds true for the number of local operations. This result is of importance for small scale implementations where the number of qubits which can stored within a device is limited. 

\begin{table}[h!]
\begin{tabular}{l|l|c|c|c}
$n$ & & D & A & B  \\
\hline & Qubits & $20$ & $14$ & $15$ \\
$5$ & CNOT & $15$ & $9$ & - \\
& Meas. & $15$ & $9$ & $10$ \\
\hline & Qubits & $90$ & $54$ & $55$ \\
$10$ & CNOT & $80$ & $44$ & - \\
& Meas. & $80$ & $44$ & $45$ \\
\hline & Qubits & $210$ & $119$ & $120$ \\
$15$ & CNOT & $195$ & $104$ & - \\
& Meas. & $195$ & $104$ & $105$
\end{tabular}
\caption{\label{tab:local:values} The table summarizes the results for the different architectures for $n=5,10,15$.}
\end{table}

\subsection{Network level}

On the network level we compare the number of qubits for the network state to connect all clients between devices. 

For the GHZ network state of architecture A this means that we account for the network proxies, not the adapater qubits, see Sec. \ref{sec:arch:ghz:net}. We elaborate our reasoning why this suffices briefly. 

We determine the number of qubits of the expanded network state (\ref{eq:network:ghz:nistatenew}) of architecture A as follows: The state $\ket{\mathrm{N}'_i}$ in (\ref{eq:network:ghz:nistatenew}), i.e. after expansion,  consists of
\begin{align}
c_i \left(1 + \sum\limits^{i-1}_{k=1} c_k \right)
\end{align}
qubits. From this we infer that the total number of qubits of the full network state after expansion is equal to
\begin{align}
\sum\limits^{m}_{i=2} \left[ c_i \left(1 + \sum\limits^{i-1}_{k=1} c_k \right) \right].
\end{align}

In order to determine the number of qubits of the secure network state for architecture B we first fix a particular device, for example device $i$: Observe that the partition of device $i$ contains $c_i$ network proxies. Now suppose we select a specific network proxy within that partition. The network state contains edges to all network proxies of other devices, therefore we have $\sum_{k \neq i} c_k$ edges associated with the selected network proxy. All these edges are decorated at least once. Furthermore, the edges to the network devices $i+1,\ldots,m$ get decorated twice which amounts to $\sum^m_{k=i+1} c_k$ additional qubits. Since we have in total $c_i$ network proxies in partition $i$ we find for the number of qubits for device $i$ the following expression:
\begin{align}
c_i \left(1 + \sum_{k \neq i} c_k + \sum^m_{k=i+1} c_k\right). \label{eq:decorated:networkqubits:device}
\end{align}
From (\ref{eq:decorated:networkqubits:device}) we find that in total
\begin{align}
\sum^m_{i=1}  \left[ c_i \left(1 + \sum_{k \neq i} c_k + \sum^m_{k=i+1} c_k\right) \right] \label{eq:decorated:networkqubits:device:trust}
\end{align}
qubits are necessary to generate the network state within the secure decorated architecture. Observe that if all network devices are trusted, then the term $\sum_{k \neq i} c_k$ in (\ref{eq:decorated:networkqubits:device:trust}) disappears. 

To compare our findings to the standard approach using Bell-pairs we first compute the number of Bell-pairs necessary to establish a fully connected network state. In particular assume that we have $m$ devices and that device $i$ connects $c_i$ clients. 

Observe that we need for each client connected to device $1$ exactly $\sum^m_{i=2} c_i$ Bell pairs which implies that device $1$ needs $c_1 \cdot \sum^m_{i=2} c_i$ in total. For device $2$ we consequently need $\sum^m_{i=3} c_i$ Bell-pairs in order to connect a single client of device $2$ to all other clients, see Sec. \ref{sec:arch:bellpairs}. Since $c_i$ clients connect to network device $i$ we easily find that the total number of Bell-pairs necessary to establish the full network state is given by
\begin{align}
\sum^{m-1}_{i=1} c_i \sum^{m}_{j=i+1} c_j.
\end{align}
We summarize our findings within Table \ref{tab:network}. Note that we consider a simpler, non-secure version of the decorated architecture with only one decorating qubit per edge for this comparison as we compare it to the standard GHZ architecture without the additional decorating qubits that guarantee security in untrusted networks.
\begin{table}[h!]
\begin{tabular}{l|c|c}
& D & A  \\
\hline Qubits & $2 \sum\limits^{m-1}_{i=1} c_i \sum\limits^{m}_{j=i+1} c_j$ & $\sum\limits^{m}_{i=2} \left[ c_i \left(1 + \sum\limits^{i-1}_{k=1} c_k \right) \right]$ \\
\hline & B & \\
\hline Qubits & $\sum\limits^m_{i=1}  \left[ c_i \left(1 + \sum\limits^m_{k=i+1} c_k\right) \right]$  & \\
\end{tabular}
\caption{\label{tab:network} The table summarizes the number of qubits occupied by the network state for the respective architectures.}
\end{table}

We emphasize that we only compare the number of qubits of the network state, not including the adapter states due to the following reasoning: the adapter states have to be generated in both, GHZ and bipartite approaches. In contrast, in case of a decorated architecture the adapter states are {\it not} necessary. 

\begin{table}[h!]
\begin{tabular}{l|l|c|c|c}
 $c$ & $m$ & D & A & B  \\
\hline & $m=5$ & $180$ & $102$ & $105$ \\ 
 $c=3$ & $m=10$ & $810$ & $432$ & $435$ \\ 
 & $m=15$ & $1890$ & $987$ & $990$ \\ 
\hline & $m=5$ & $500$ & $270$ & $275$ \\ 
 $c=5$ & $m=10$ & $2250$ & $1170$ & $1175$ \\ 
 & $m=15$ & $5250$ & $2695$ & $2700$ \\ 
 \hline & $m=5$ & $980$ & $518$ & $525$ \\ 
 $c=7$ & $m=10$ & $4410$ & $2268$ & $2275$ \\ 
 & $m=15$ & $10290$ & $5243$ & $5250$ 
\end{tabular}
\caption{\label{tab:network:values} The table summarizes the number of qubits which need to be stored according to Table \ref{tab:network} for the corresponding architectures in different settings of $m$ and $c_i = c$ for all $i$.}
\end{table}
From Table \ref{tab:network:values} we find: our multipartite approaches outperform the bipartite approach via Bell-pairs. Furthermore we observe that the decorated network state may occupy slightly more qubits than the GHZ network state, but this state does not require any adapter states.

\section{Optimized architectures}\label{sec:optimized}

We also want to emphasize if less functionality, or equivalently the set of target graph states is more restricted, then our architectures need far less states/qubits. Therefore, in a restricted setting, our architecture performs far more efficient compared to bipartite approaches. 

For example consider the scenario of a network which is restricted to generate GHZ states only. Such a network might be of interest for a community or group of clients which want to agree on conference keys via quantum key agreement protocols relying on GHZ states, but also for distributed sensing or other tasks. 

If only GHZ states should be provided by a network with $m$ devices, then it suffices to generate the network state
\begin{align}
\ket{\mathrm{N}} = \ket{\mathrm{GHZ}_{m}}\label{eq:ghz:optimized}
\end{align}
instead of (\ref{eq:network:ghz:nstate}) and (\ref{eq:network:ghz:nistate}). Furthermore, we replace the device state for devices using GHZ states of (\ref{eq:ghz:newdevice}) with
\begin{align}
\ket{\mathrm{D}_i} = \ket{\mathrm{GHZ}_{c_i + 1}} \label{eq:ghz:optimized:device}
\end{align}
where $1 \leq i \leq m$. We observe that via the states of (\ref{eq:ghz:optimized}) and (\ref{eq:ghz:optimized:device}) the network is able to create any GHZ state in the network. 

If the network should also account for parallel requests in that setting, then we enhance the network and devices as follows: instead of generating a single copy of (\ref{eq:ghz:optimized}) and (\ref{eq:ghz:optimized:device}) the network and the devices prepare $\nu$ copies of both states. This enables the network to work on $\nu$ requests in parallel. 

One may optimize our architectures in a similar fashion also to other classes of target states, like e.g. open or closed cluster states. 

Another optimization technique of the architecture of Sec. \ref{sec:arch:ghz} is to store a modified version of the network state of Sec. \ref{sec:arch:ghz:net}. In particular, by symmetrizing the GHZ states of (\ref{eq:network:ghz:nstate}), i.e. distributing several copies of that state where the roots of the GHZ states get shifted among the devices, one can minimize the number of qubits which need to be measured within the $Z$ basis for particular requests. For example, a network device might choose that GHZ state of the network which matches the number of neighbouring devices for a particular request. One could even further optimize such an architecture by shifting around the root of the GHZ network state to different network devices as those other graph states are Local-Clifford equivalent to the GHZ network state (via local complementation, see e.g. \cite{He04, He06}).

\section{Conclusion and outlook}\label{sec:conclusion}

We have presented elementary buildings blocks at an abstract level for quantum networks which enable the generation of arbitrary graph states in a highly distributed manner. In particular we identified two architectures: an architecture relying on GHZ states and an architecture which uses decorated graph states. Clients solely connect to quantum network devices and both architectures make extensive use of multipartite entangled quantum states. Our approaches are plug-and-play, which means clients and network devices can appear or also disappear at any time within the network. This enables for dynamic networking. The device and network state within GHZ architectures corresponds to several GHZ states of different sizes. In contrast, the states within the decorated architecture use a $m-$partite or fully connected, decorated graph state. In addition, for each architecture we have presented a protocol for generating, distributing and extending the network state. Our approaches turn out to be especially suited for recursive networks. Finally we have shown the security of our architectures in different settings, ranging from trusted networks to completely untrusted network where only the local apparatus of each client is trustworthy. 

What remains to be dealt with are imperfections in the apparatus of the network devices. Therefore one can consider to employ entanglement distillations protocols to cope with noise and imperfections. In particular it would be interesting to investigate specific noise models and entanglement distillation protocols to obtain error thresholds, reachable fidelities and concrete security levels of our protocols in a noisy environment. In addition, it is not clear which architecture is best suited for a concrete noisy network where the number of devices and number of clients connected to a device might vary. Furthermore, there might also exist other multi-partite entangled states besides GHZ states and decorated, fully connected graph states which could be candidates for network states. It is also not clear if the states we use are optimal in terms of storage size. Finally, it would be interesting to study how classical information needs to propagate within our network architectures.


\section*{Acknowledgements.---} This work was supported by the Austrian Science Fund (FWF) through projects P28000-N27, P30937-N27 and SFB F40-FoQus F4012-N16.

\appendix

\section{GHZ architecture example \label{app:example_ghz}}

In this section we describe an example request in a network consisting of $3$ devices with $c_i = 3$ clients each in order to illustrate the workings of the protocol in detail.

\begin{figure}
 \includegraphics[width=0.8\columnwidth]{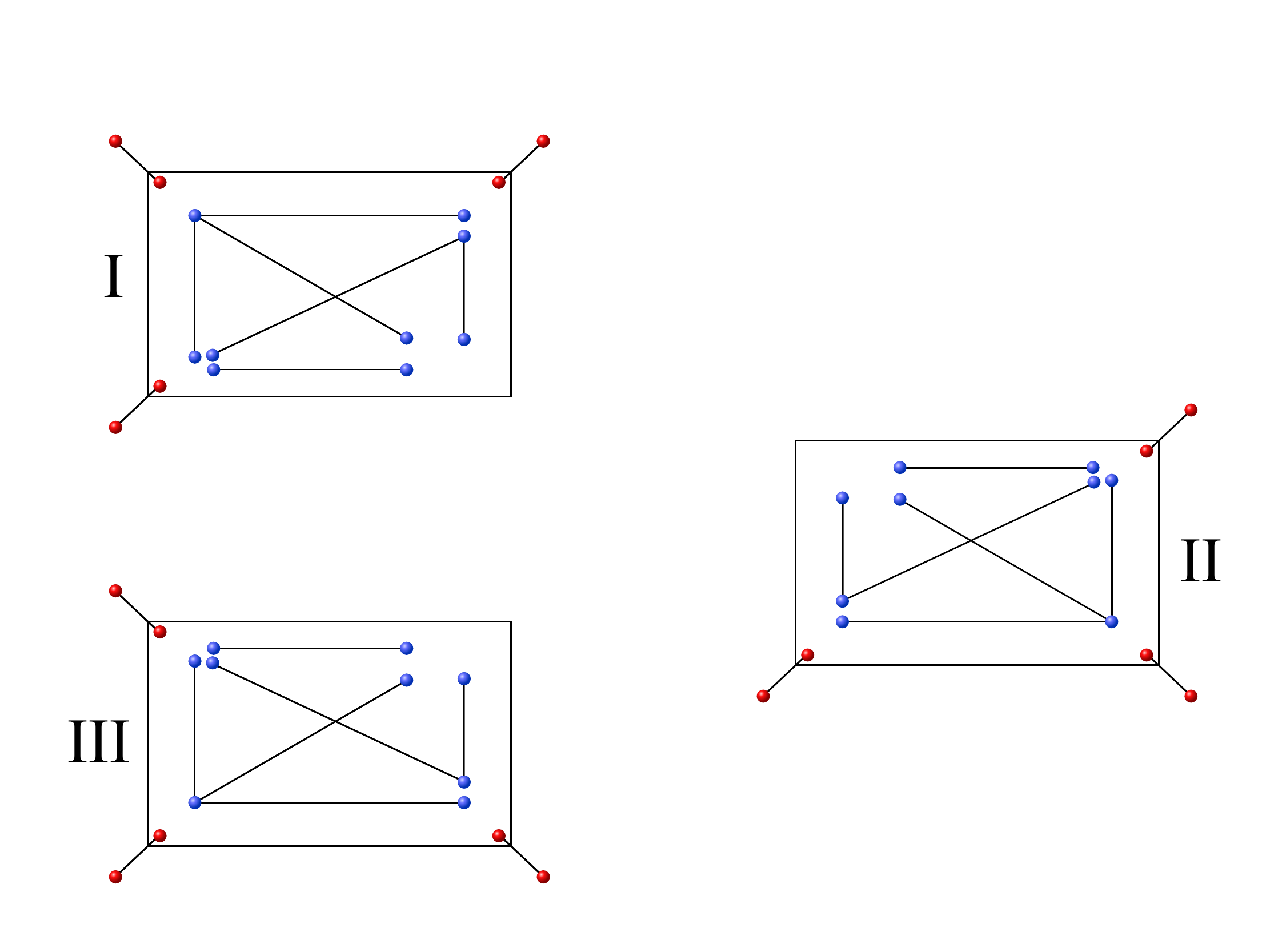}
 \caption{Preparation of the device states. \label{fig:example_ghz_step1}}
\end{figure}

First, the states that will be the resource for the network need to be generated. Each of the devices prepares the device states for four parties to connect their three clients and the connection to the outside network. In this specific example the states per device are one $4$-qubit GHZ state, one $3$-qubit GHZ state and a $2$-qubit GHZ state (Bell pair), as depicted in Fig. \ref{fig:example_ghz_step1}.

\begin{figure}
 \includegraphics[width=0.8\columnwidth]{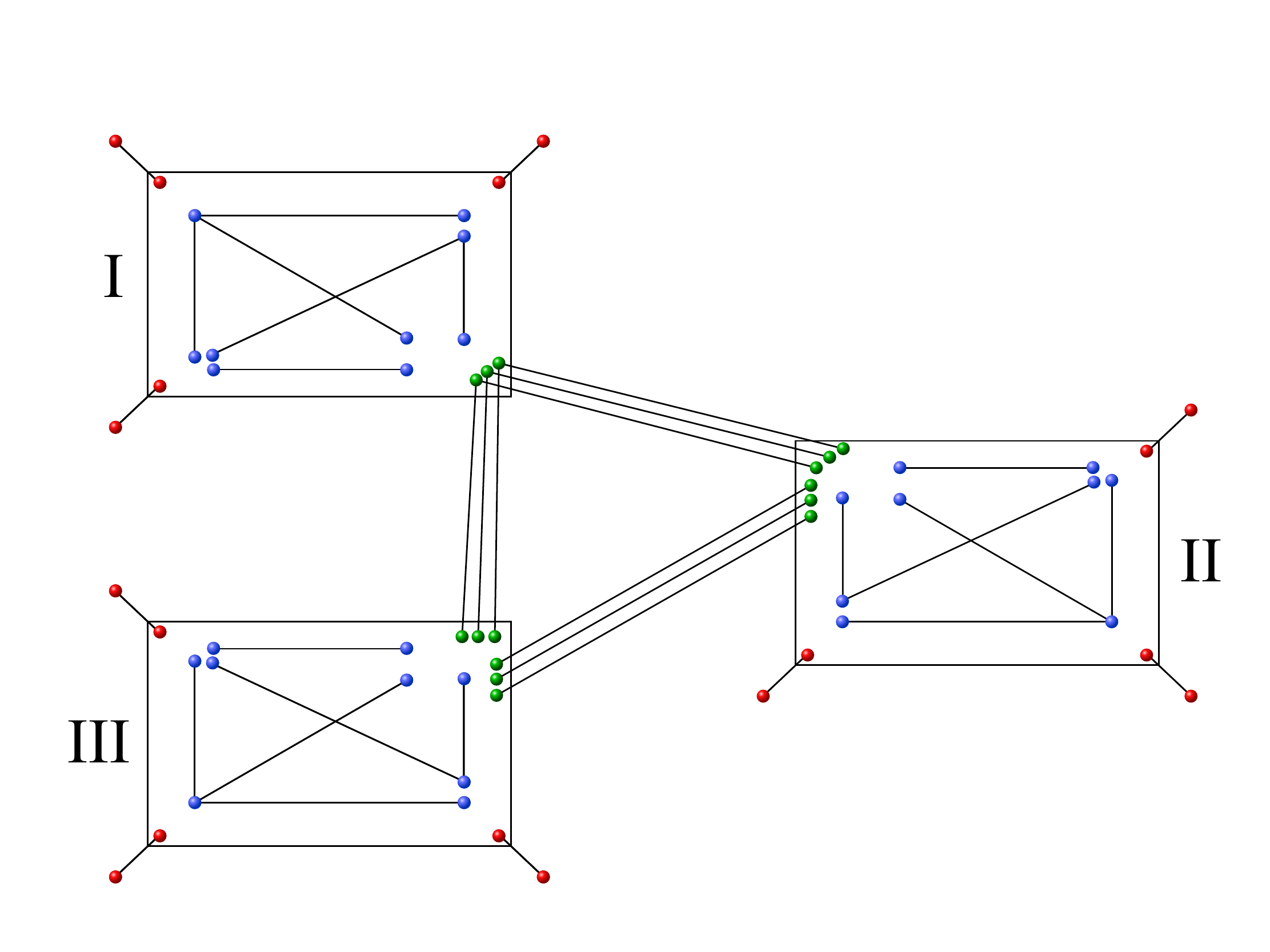}
 \caption{The network states are distributed. The network is now ready to receive requests. \label{fig:example_ghz_step2}}
\end{figure}

Then, the network states are distributed between the devices in the network, which can be seen in Fig. \ref{fig:example_ghz_step2}. Take note that three copies of the GHZ state and three copies of the Bell state are needed to ensure full functionality, i.e. any graph state can be distributed between the clients of the three network devices.

After this step the network is ready to receive requests from the clients.

\begin{figure}
 \includegraphics[width=0.8\columnwidth]{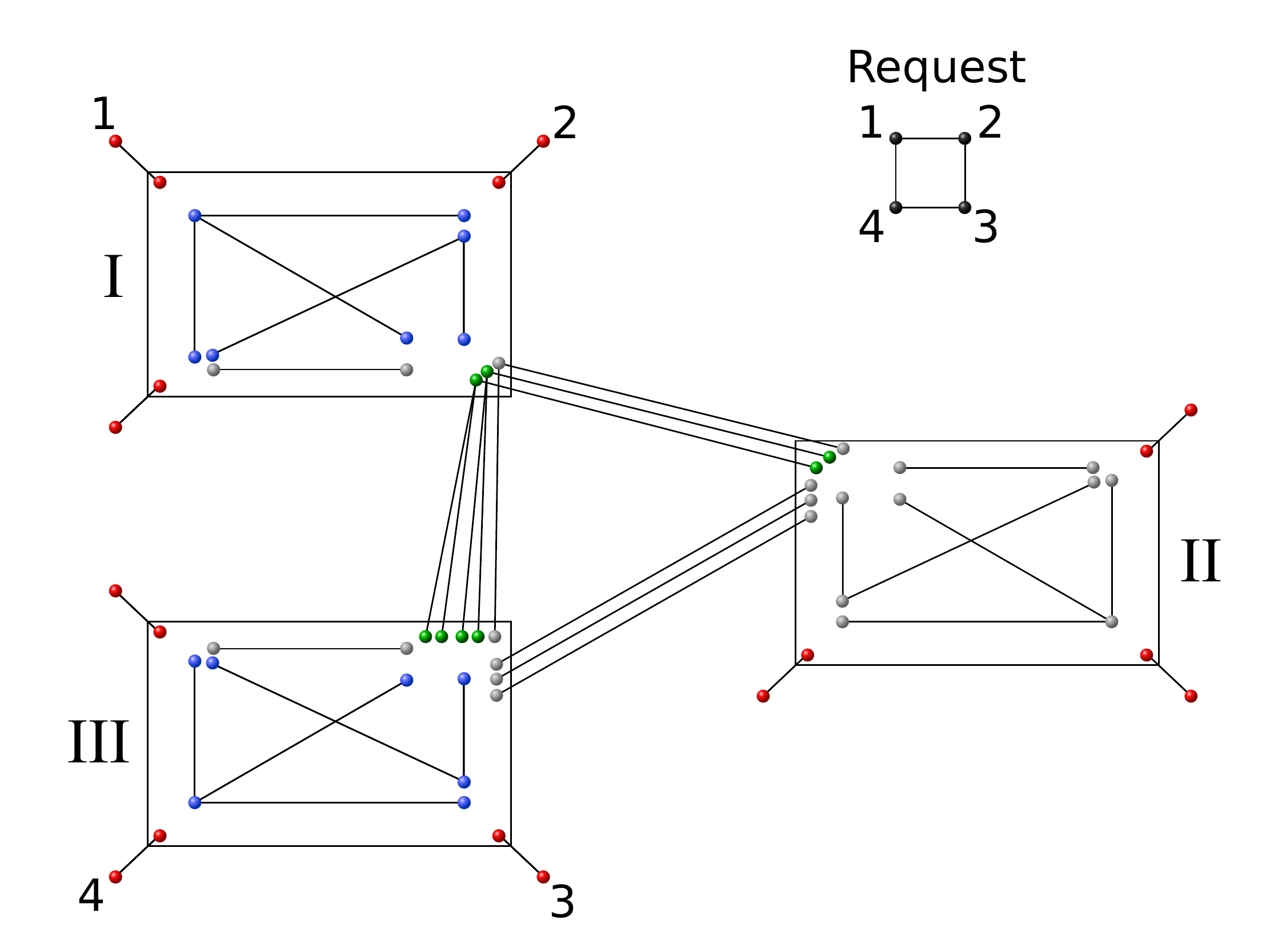}
 \caption{The network receives the request to distribute a particular graph state between the clients marked $1$ to $4$. The states which are not needed to fulfill the request (greyed out) can be saved for future requests. The GHZ states are expanded using expander states according to how many clients in the device are part of the requested state. \label{fig:example_ghz_step3}}
\end{figure}

Let us look at a particular request of a graph state corresponding to a $4$-qubit ring graph between the clients marked $1$ to $4$ in Fig. \ref{fig:example_ghz_step3}. The network can identify which states are needed to fulfill this request using only the information about how many clients are part of the target state for each device. In this particular example only two of the $3$-qubit GHZ states are needed because the request involves only parties connected to devices I and III and $2$ clients are connected to the device with the root of the GHZ state. Note that the states which are not used for this request can be stored for later use. Then, the devices with the leafs of the resource states use expander states to expand the GHZ states according to how many clients are part of the requested state in that particular device. Of course, each device can simply fully expand the states to the number of clients which are connected to that device. However, this is not necessary if only some clients are involved in the request. In Fig. \ref{fig:example_ghz_step3} the situation after the expansion is shown.

\begin{figure}
 \includegraphics[width=0.8\columnwidth]{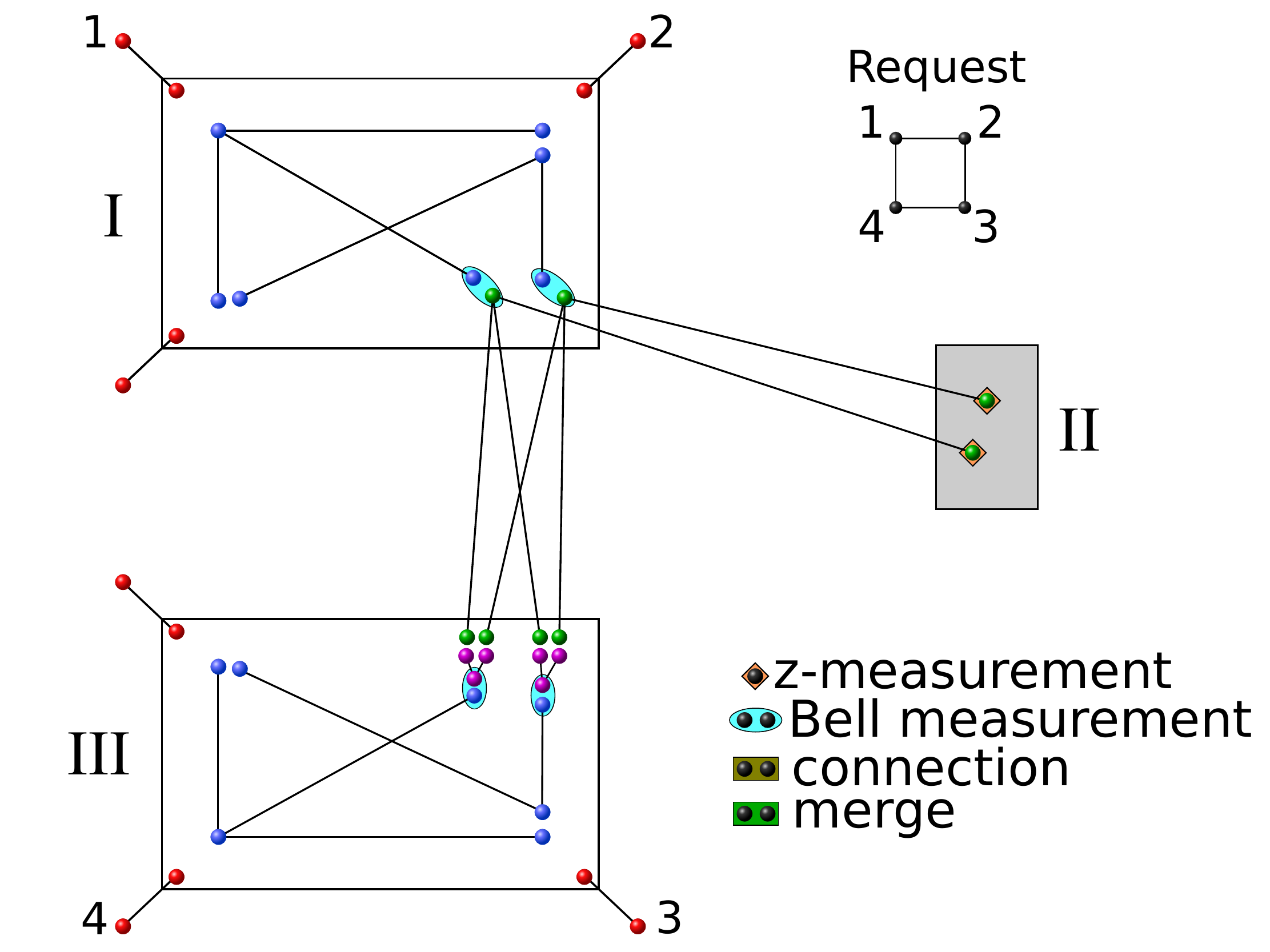}
 \caption{Generation of adapter states and connection to device states. \label{fig:example_ghz_step4}}
\end{figure}

Next, the devices generate the adapter states, which are GHZ states as well, that will be needed to pick the desired adjacencies in the graph state. The roots of the network states and the adapter states are then connected to the device states via Bell measurements as illustrated in Fig. \ref{fig:example_ghz_step4}. Device II is not involved in this particular request and will simply measure its qubits belonging to the network state in the $Z$ basis. Since device II is not participating in this case, its device state is not used up and can be used to fulfill parallel requests that only involve clients connected to device II. Here it also becomes obvious why it is necessary to add decorating qubits as a way to reliably cut the connections to device II (as described in section \ref{sec:arch:ghz:sec}), if that device, which is not directly involved in the generation of the requested state, is not trusted.

\begin{figure}
 \includegraphics[width=0.8\columnwidth]{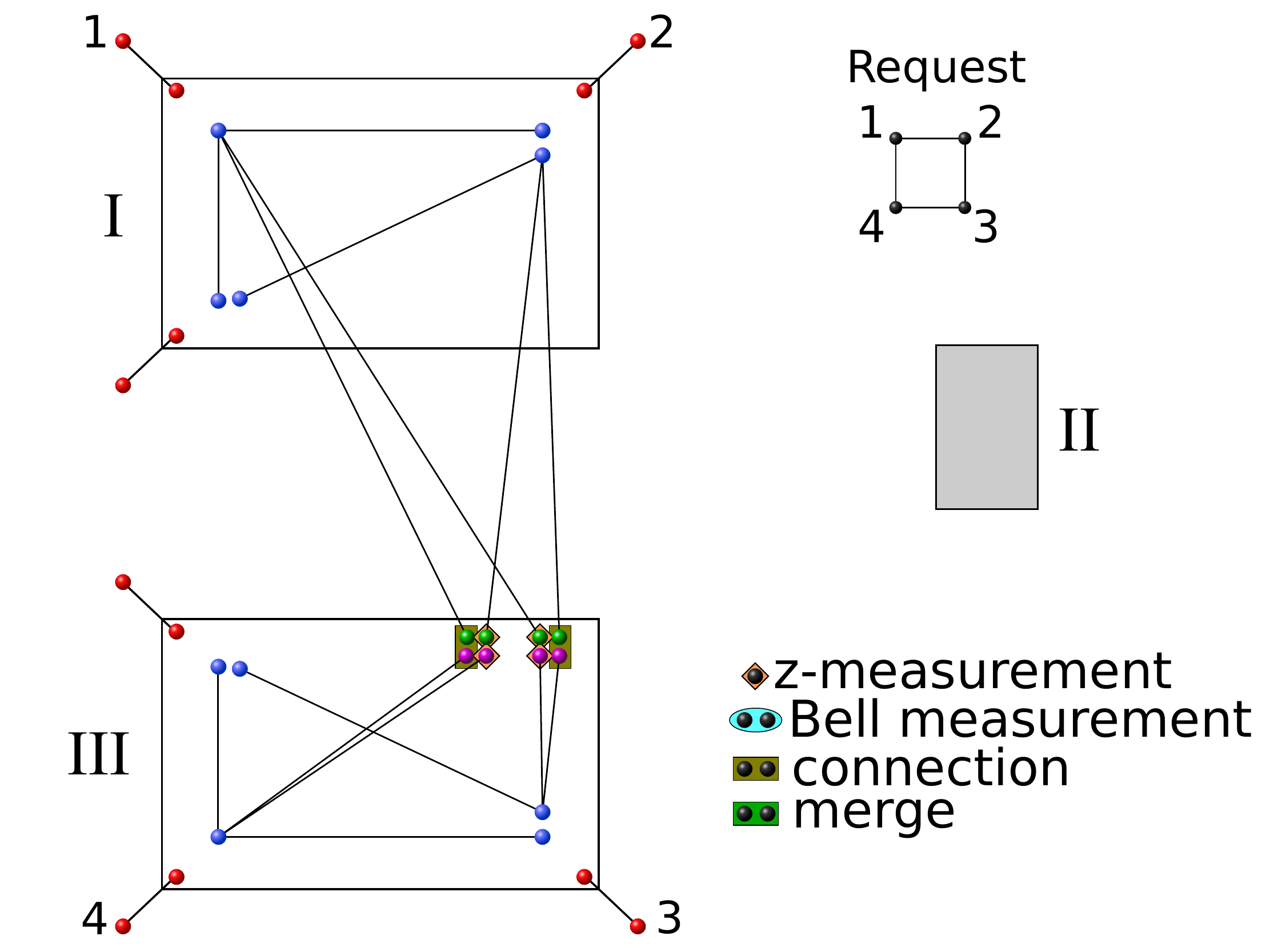}
 \caption{The adjacencies at the network level are generated. \label{fig:example_ghz_step5}}
\end{figure}

The remaining qubits, i.e. the leafs of the expander and adapter states corresponding to the network level, are then used to generate the desired adjacencies in the graph states. To establish an edge in the graph, the qubits are connected using the connect procedure for graph states of Sec. \ref{sec:back:merge:czconnect}, i.e. a controlled phase gate followed by $Y$ measurements (and local Clifford corrections). Otherwise, the connecting qubits are simply measured in the $Z$ basis to remove them from the graph. This process is shown for our example state in Fig. \ref{fig:example_ghz_step5}.

\begin{figure}
 \includegraphics[width=0.8\columnwidth]{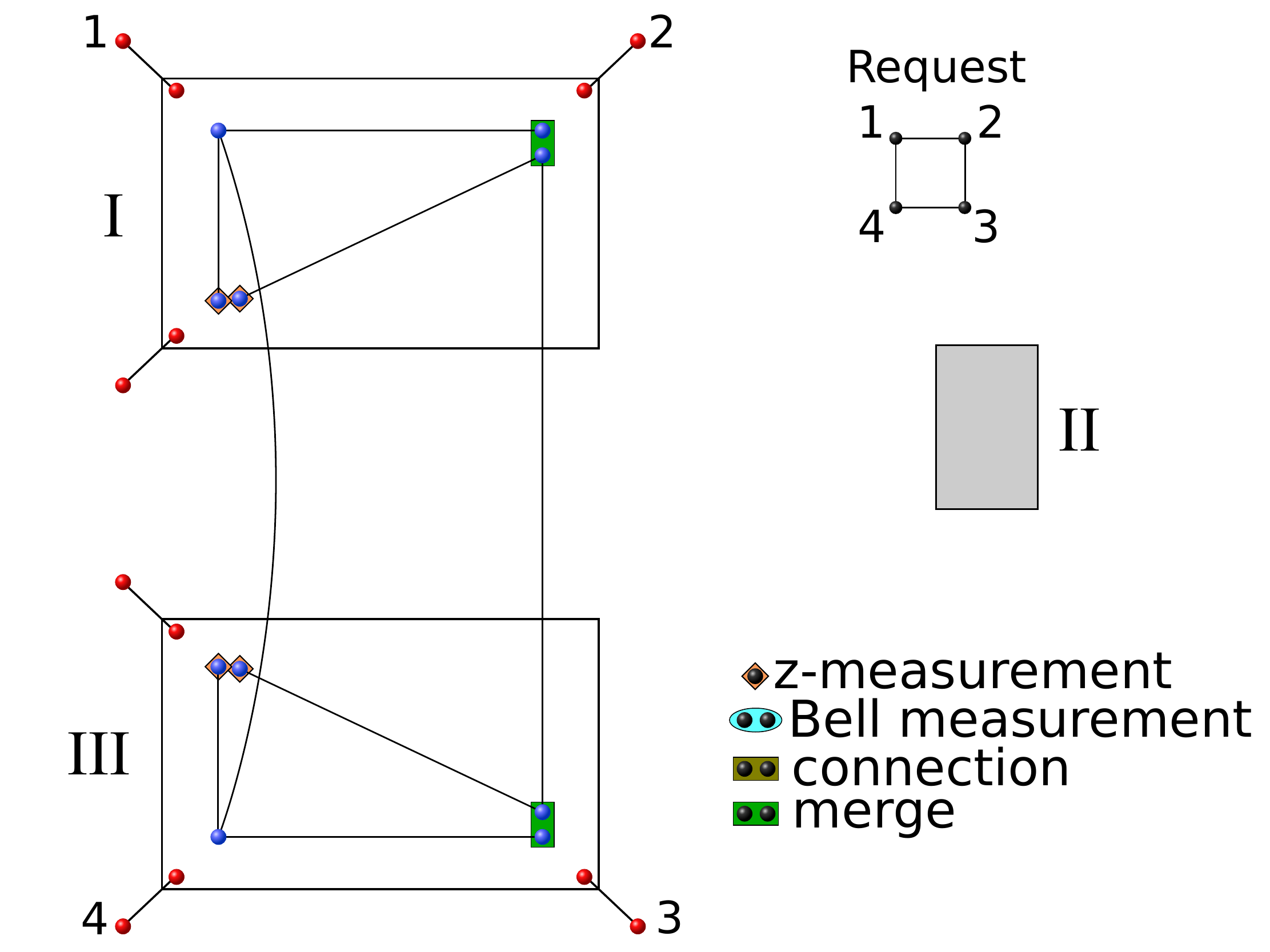}
 \caption{The adjacencies at the device level are generated. \label{fig:example_ghz_step6}}
\end{figure}

Similarly, the adjacencies between the parties directly connected to a device are established. In particular, unwanted connections are removed by measuring directly in the $Z$ basis while desired connections are established via the merging procedure of Sec. \ref{sec:back:merge:cnot}, i.e. a controlled NOT gate and a $Z$ measurements, see Fig. \ref{fig:example_ghz_step6}.

\begin{figure}
 \includegraphics[width=0.8\columnwidth]{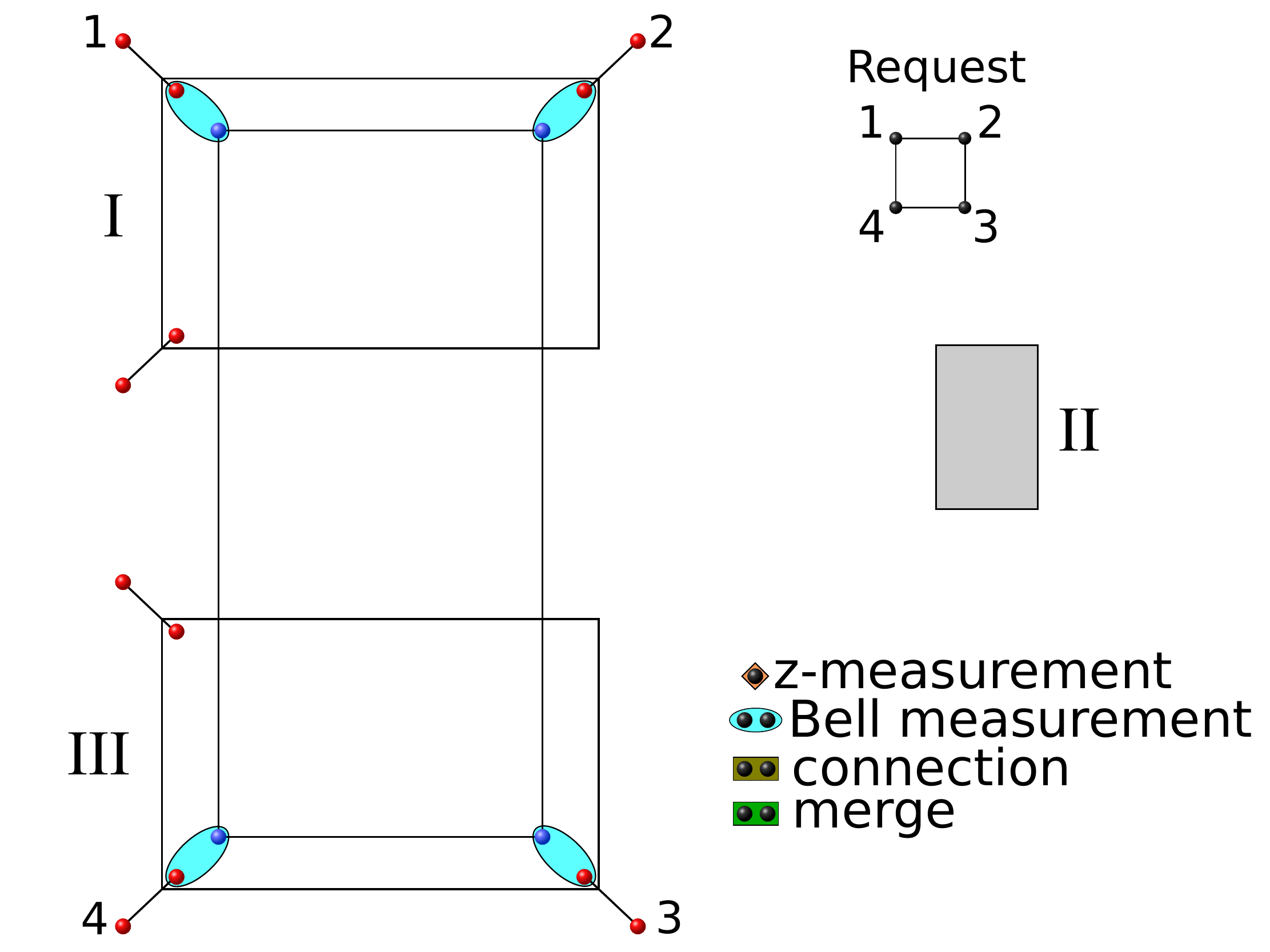}
 \caption{The final graph state is teleported to the clients and the request was successfully fulfilled.  \label{fig:example_ghz_step7}}
\end{figure}

Finally, the finished graph state is teleported to the clients, see Fig. \ref{fig:example_ghz_step7}.

\section{Parallel request example \label{app:example_parallel_request}}
One useful aspect of the GHZ network architecture is that the network can handle parallel requests between disjoint sets of parties. In the above example, in Fig. \ref{fig:example_ghz_step3} the qubits of several states are greyed out because those states are not used for fulfilling the particular request discussed there. These left-over states can be used to fulfill another request in parallel.

\begin{figure}
 \includegraphics[width=0.8\columnwidth]{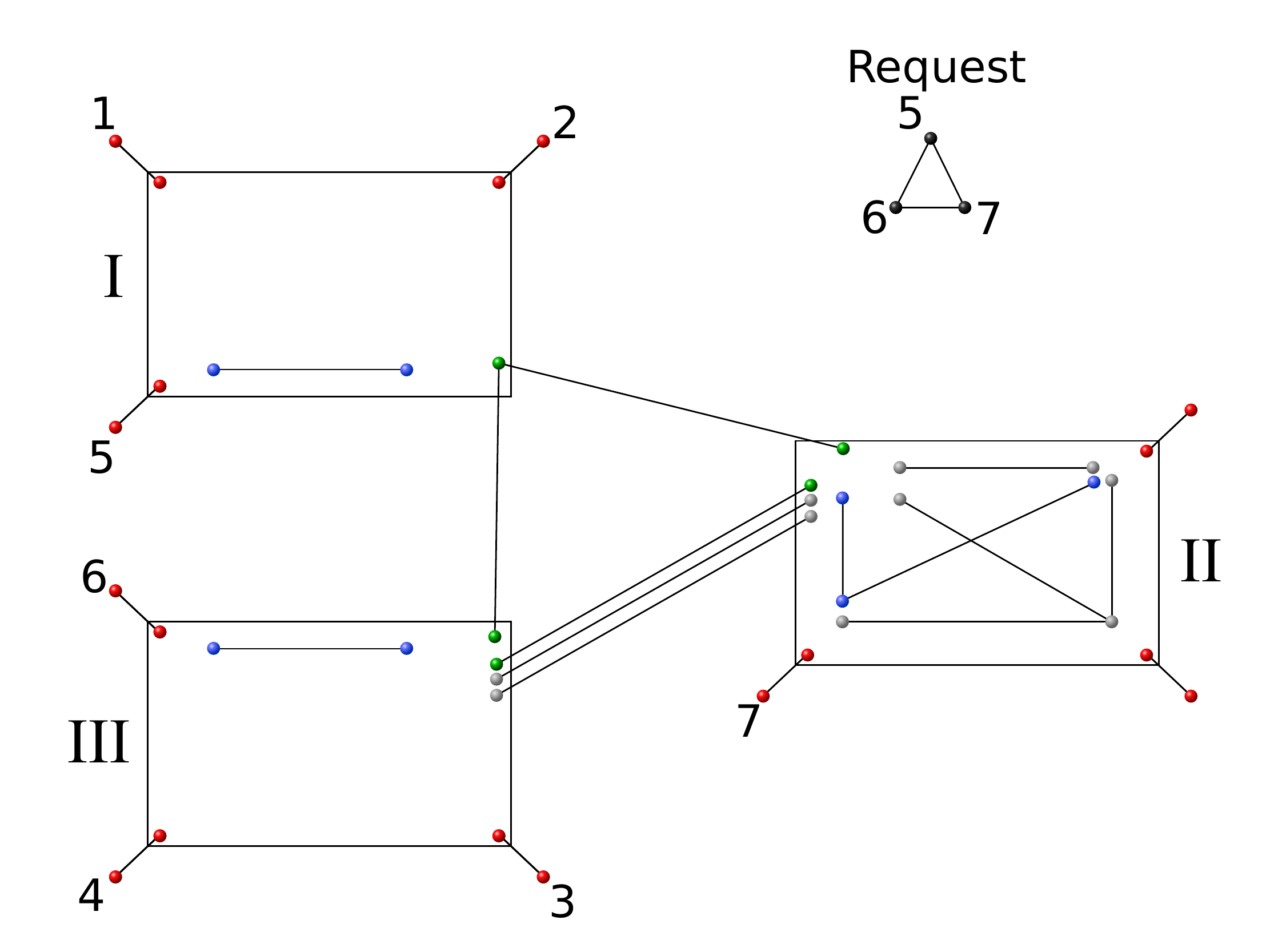}
 \caption{ The network receives the parallel request to distribute a graph state between parties $5$, $6$ and $7$. Only the states not used up by the other request are shown.\label{fig:example_parallel_request_step1}}
\end{figure}

Let us consider a parallel request that consists of distributing the fully connected graph state between the parties labelled 5-7 in Fig. \ref{fig:example_parallel_request_step1}, which also depicts all the states that are not in use for the other request.

\begin{figure}
 \includegraphics[width=0.8\columnwidth]{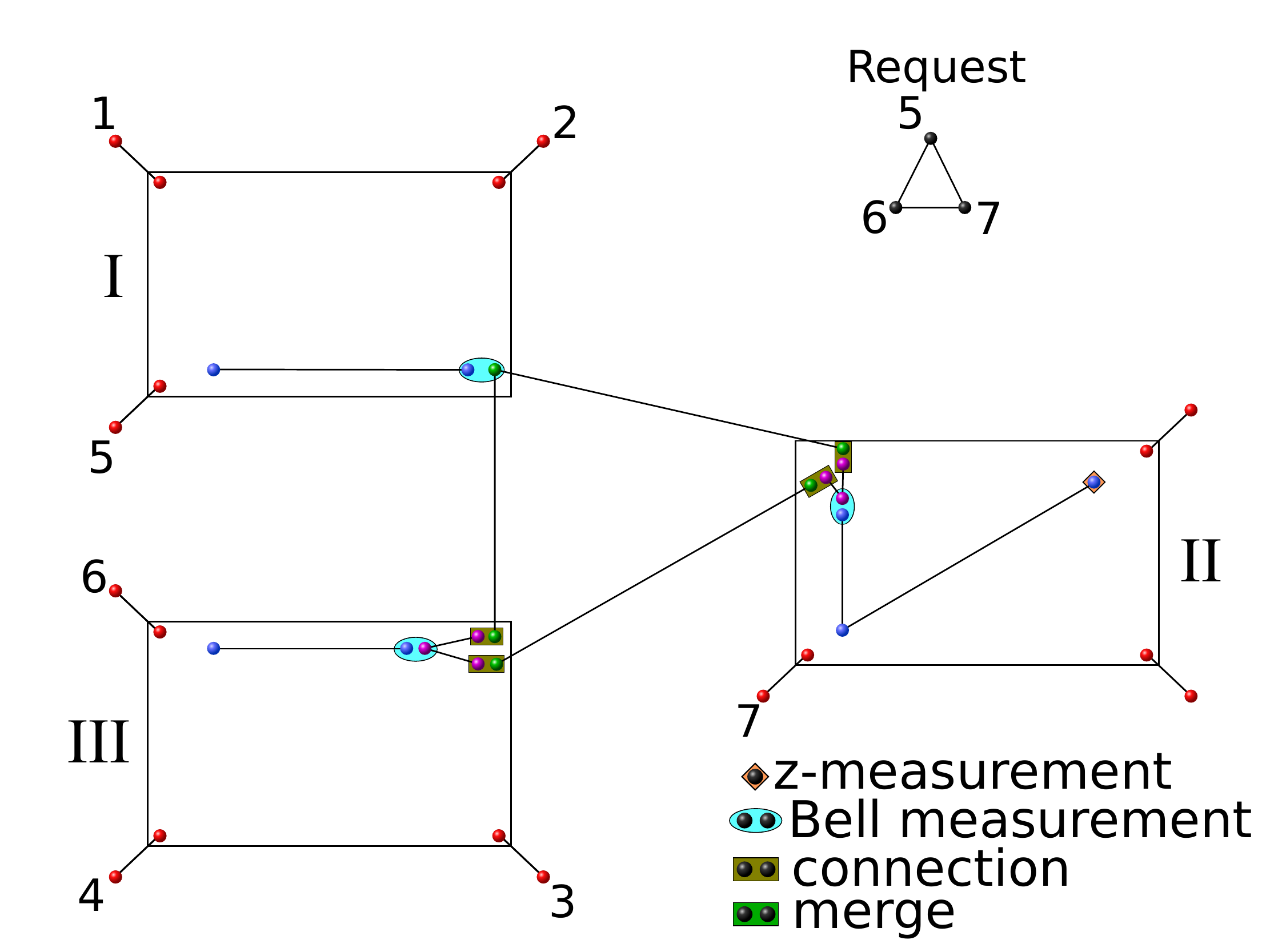}
 \caption{ Using only states not used by the other request, the network can fulfill this parallel request by applying the appropriate measurements.\label{fig:example_parallel_request_step2}}
\end{figure}

From here on, we just follow the usual protocol for generating the graph state. Note that in this particular case there are no expander states needed because only one party per device is participating in this request. The operations and measurements as specified in Fig. \ref{fig:example_parallel_request_step2} are applied and all that is left to do is to teleport the state to the clients.

\bibliographystyle{apsrev4-1}
\bibliography{quantum_networks}

\end{document}